\documentclass[reprint, PRL, APS]{revtex4-1}
\usepackage[english]{babel}
\usepackage{graphicx}
\usepackage[bbgreekl]{mathbbol}
\usepackage{amsfonts}
\usepackage{amsmath}
\usepackage{hyperref}
\usepackage{cancel}
\usepackage{colortbl}
\usepackage{rotating}
\usepackage{siunitx}
\DeclareSymbolFontAlphabet{\mathbb}{AMSb}
\DeclareSymbolFontAlphabet{\mathbbl}{bbold}

\newcommand{\blue}[1]{\textcolor{blue}{#1}}
\newcommand{\kbt}{k_{ B}T}
\newcommand{\nn}{\nonumber\\}

\begin{abstract}
A detailed comparison is made between different viewpoints on reversible heating in electric double layer capacitors. 
We show in the limit of slow charging that a combined Poisson-Nernst-Planck and heat equation, first studied by d'Entremont and Pilon [J. Power Sources {\bf 246}, 887 (2014)], recovers the temperature changes as predicted by the thermodynamic identity of Janssen \textit{et al.} [Phys. Rev. Lett. {\bf 113}, 268501 (2014)], and disagrees with the approximative model of Schiffer \textit{et al.}  [J. Power Sources {\bf 160}, 765 (2006)] that predominates the literature. The thermal response to the adiabatic charging of supercapacitors contains information on electric double layer formation that has remained largely unexplored. 
\end{abstract}

\begin{document}

\author{Mathijs Janssen and Ren\'{e} van Roij}
\affiliation{Institute for Theoretical Physics, Center for Extreme Matter and Emergent Phenomena,  Utrecht University, Princetonplein 5, 3584 CC Utrecht, The Netherlands}
%
%
%
%
%
\date{\today}
\title{Reversible heating in electric double layer capacitors}
\maketitle
With the relation between heat and entropy formulated by Clausius in 1855, and with the establishment of the importance of ion entropy to the electric double layer (EDL) by Gouy (1910) and Chapman (1913) \cite{gouy1910constitution}, almost a century passed before reversible, adiabatic heating and cooling was measured in electric double layer capacitors (EDLCs) \cite{schiffer2006heat}. Unlike irreversible Joule heating, occurring everywhere in the electrolyte when an EDLC is charged at finite currents, it turns out that the sources of reversible heating are located only within the nanometer-range vicinity of the electrode's surface. Therefore, one needs EDLCs whose surface-to-volume ratio is as high as possible to notice an appreciable reversible temperature variation. This has become possible and relevant in recent years because electrodes can now be manufactured from porous carbon with internal surface areas up to 2000~m$^2$g$^{-1}$. Electrolyte-filled supercapacitors made from these electrodes are characterized by a high capacitance, fast (dis)charging rates, and high cyclability \cite{simon2008materials}. These favorable properties have sparked a huge scientific interest in supercapacitors in recent years, and led to various applications \cite{brogioli2009extracting, Janssen:2014aa, hartel2015heat, hamelers2013harvesting,suss2015water}. 
The performance of supercapacitors for energy storage usually suffers, however, from increased temperatures causing aging of materials, increased internal resistance, decreased capacitance, parasitic electrochemical reactions, and self discharging \cite{xiong2015thermal, miller2006electrochemical, zhang2016reversible}. Efforts were therefore made both in experiments  \cite{schiffer2006heat, miller2006electrochemical, pascot2010calorimetric, zhang2016reversible, gualous2011supercapacitor} and modeling \cite{d2014first, lee2014modeling,d2014first2,kumar2015microscopic} to gain insight in the thermal behavior of supercapacitors.
However, a unified understanding of reversible heating effects occurring during EDL buildup is still lacking, and thermal response to charging has not yet been fully exploited.
This Letter for the first time quantitatively reconciles two viewpoints on reversible heating. Within the {\it thermodynamic} viewpoint, two distinct identities for isentropic processes are discussed, only one of which (we show) agrees with the other, {\it kinetic}, viewpoint. 
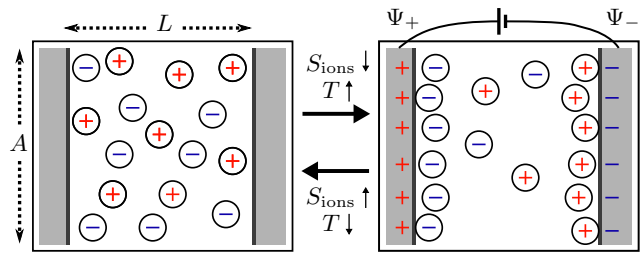
\begin{figure}[b]
\centering
\def\svgwidth{0.475\textwidth}{ 
  \providecommand\transparent[1]{%
    \errmessage{(Inkscape) Transparency is used (non-zero) for the text in Inkscape, but the package 'transparent.sty' is not loaded}%
    \renewcommand\transparent[1]{}%
  }%
  \providecommand\rotatebox[2]{#2}%
  \ifx\svgwidth\undefined%
    \setlength{\unitlength}{1539.3125bp}%
    \ifx\svgscale\undefined%
      \relax%
    \else%
      \setlength{\unitlength}{\unitlength * \real{\svgscale}}%
    \fi%
  \else%
    \setlength{\unitlength}{\svgwidth}%
  \fi%
  \global\let\svgwidth\undefined%
  \global\let\svgscale\undefined%
  \makeatother%
  \begin{picture}(1,0.38091681)%
    \put(0,0){\includegraphics[width=\unitlength]{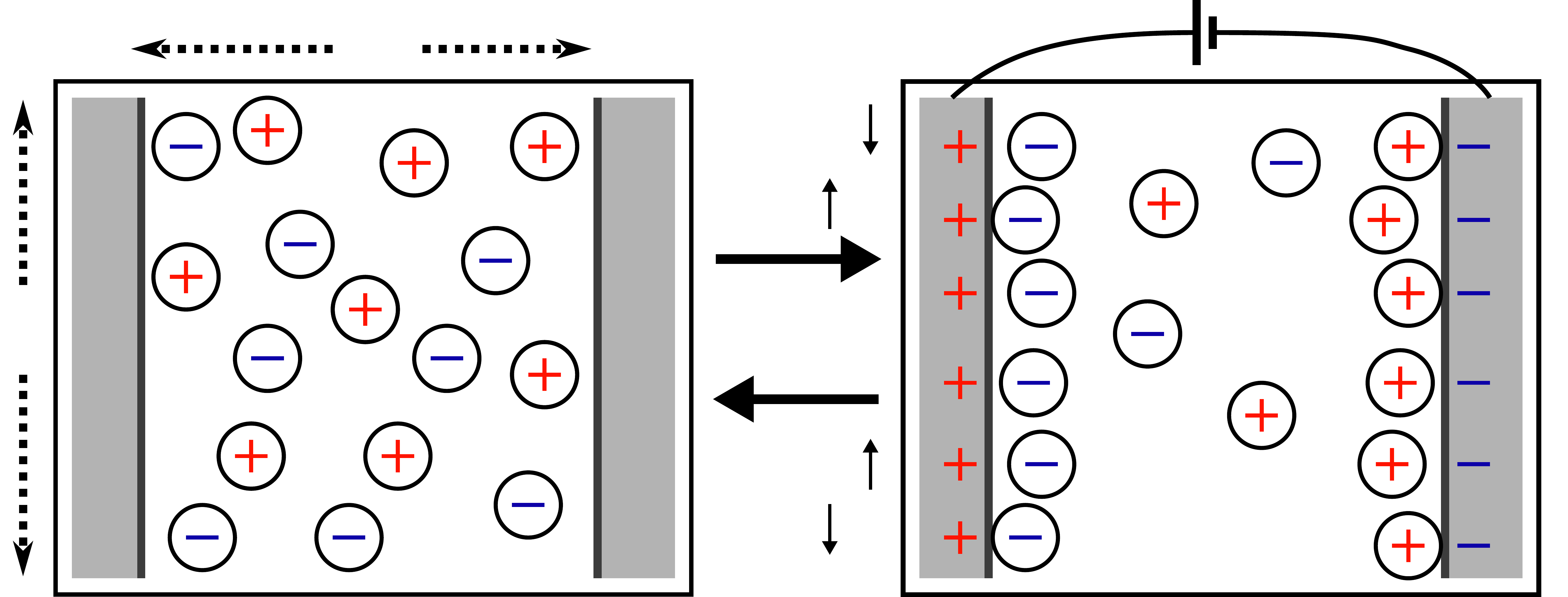}}%
    \put(0.46166714,0.28321751){\color[rgb]{0,0,0}\makebox(0,0)[lb]{\smash{$S_{\rm ions}$}}}%
    \put(0.48765277,0.24060108){\color[rgb]{0,0,0}\makebox(0,0)[lb]{\smash{$T$}}}%
    \put(0.46166714,0.07429307){\color[rgb]{0,0,0}\makebox(0,0)[lb]{\smash{$S_{\rm ions}$}}}%
    \put(0.48765277,0.03271607){\color[rgb]{0,0,0}\makebox(0,0)[lb]{\smash{$T$}}}%
    \put(0.2277965,0.34454359){\color[rgb]{0,0,0}\makebox(0,0)[lb]{\smash{$L$}}}%
    \put(-0.00087701,0.15744708){\color[rgb]{0,0,0}\makebox(0,0)[lb]{\smash{$A$}}}%
    \put(0.58639815,0.35493784){\color[rgb]{0,0,0}\makebox(0,0)[lb]{\smash{$\Psi_{+}$}}}%
    \put(0.92940842,0.35493784){\color[rgb]{0,0,0}\makebox(0,0)[lb]{\smash{$\Psi_{-}$}}}%
  \end{picture}%
}
\caption{Slow charging of a thermally insulated EDLC reduces the ionic configuration entropy $S_{\rm ions}$ and thereby causes a rise in the temperature $T$ of the electrolyte (solvent not shown). Upon discharging an opposite cooling effect is observed.}
\label{fig1}
\end{figure} 

For the thermodynamic viewpoint, consider an EDLC on which a potential is imposed by connecting it to a battery. The electrodes then obtain surface charges which are screened by diffuse clouds of counterionic charge (see Fig.~\ref{fig1}), hence the ionic configuration entropy decreases. For a thermally insulated capacitor that is charged quasistatically, thermodynamics demands via the second law ($dS=0$) that this decrease is counterbalanced by an electrolyte entropy increase: the EDLC heats up. Upon quasistatic adiabatic discharging the opposite happens: while the EDL breaks down the electrolyte cools. Experimental observations of reversible heating in an EDLC were first reported in Ref.~\cite{schiffer2006heat} (and later in Refs.~\cite{pascot2010calorimetric, gualous2011supercapacitor}). Here, the EDL buildup was described theoretically as an isentropic compression of an ideal gas. While this model correctly captures the exchange between configuration and momentum contributions to the fixed phase space volume, it completely ignores the long-range Coulomb interactions among the constituent particles. An alternative expression not hinging on ideal-gas reasoning was proposed by the current authors in Ref.~\cite{Janssen:2014aa} [and repeated here in Eq.~(\ref{eq:2adiabat})]. Interestingly, there are many well-established examples of isentropic temperature changes which are governed by equations analogous to Eq.~(\ref{eq:2adiabat}), e.g., the magnetocaloric \cite{pecharsky1999magnetocaloric},  the electrocaloric \cite{mischenko2006giant}, and the Joule-Gouge effect  \cite{schweizer2001atomistic}. 

The kinetic viewpoint on heat production in the EDL can be traced back  to Verwey and Overbeek \cite{verwey1948theory} who stated for EDL discharging that 
``..the counter ions must diffuse more and more back into the solution. This diffusion occurs against electric forces... 
The energy needed to raise the electric energy of these ions must be taken up from the surrounding ions and molecules, and is delivered as kinetic energy from the thermal motion of the latter.''  
A thermally insulated EDLC therefore cools upon discharging.  This exchange between electric energy and heat is captured in the internal energy balance [see Eqs.~(\ref{eq:heatequation}) and (\ref{eq:energy_balance})] as the inner product ${\bf I}\cdot {\bf E}$ of the ionic current ${\bf I}$ and the electric field  ${\bf E}$ \cite{groot1962non, haase1968thermodynamics,kontturi2008ionic}. 
In bulk electrolytes ions respond Ohmically to an imposed electric field [see Eq.~(\ref{eq:ioniccurrent})]: the electric field drives a current that is proportional to and aligned with this field. Hence, ${\bf I}\cdot {\bf E}\sim  I^{2}>0$ in the bulk, such that electric energy is irreversibly transferred to the internal energy of the fluid, also known as Ohmic losses or Joule heat. 
In general, however, the direction of particle fluxes is set by the gradient of the {\it electrochemical} potential, which, next to the gradient in electric potential, also contains the gradient of the local ion density. In regions of strong concentration gradients it is therefore possible that the ionic current opposes the electric forcing (${\bf I}\cdot {\bf E}<0$), giving rise to localized cooling \cite{d2014first, kontturi2008ionic,biesheuvel2014negative}. 
Since the sources of the reversible heating are located only in the EDL, the resulting temperature variations are more pronounced in supercapacitors that have a large surface-to-volume ratio. However, while supercapacitors have a highly intricate pore structure, this Letter focuses for illustration purposes on a parallel plate EDLC  as it already captures the essential physics. Capturing ionic currents within a modified Poisson-Nernst-Planck model, we show that the heat equation recovers the prediction of Eq.~(\ref{eq:2adiabat})  in the limit of slow charging, thereby reconciling the thermodynamic and kinetic viewpoints. The ideal-gas reasoning, often used for illustration purposes \cite{xiong2015thermal, zhang2016reversible, d2014first}, or even to fit to experiments \cite{gualous2011supercapacitor, lee2014modeling}, cannot reproduce the adiabatic temperature change.  

Consider a thermally insulated container with two planar parallel electrodes separated by an (incompressible) 1:1 electrolyte of dielectric constant $\epsilon$ (Fig.~\ref{fig1}). The electrolyte consists of $2N=N_{+}+N_{-}$ ions and $N_{s}$ solvent (s) molecules, and occupies a volume $V_{\rm el}=A L$ where $L$ is the electrode separation, and $A$ the large surface area of each of the electrodes. An external battery imposes the potential $\Psi$ and $-\Psi$ to the ideally polarizable electrodes, leading to opposite surface charges $Q$ and $-Q$ with corresponding surface charge densities $\pm e\sigma=\pm Q/A$, where $e$ is the proton charge. The coordinate $z$ runs perpendicular to the plates from $z=0$ to $z=L$. At finite potentials, ionic density profiles $\rho_{\pm}(z)$ are inhomogeneous because an EDL is formed to screen the surface charge. The bulk salt concentration is defined in the uncharged state as $\rho_{0}=N/V_{\rm el}$. 
Since the dielectric constant $\epsilon(T,\rho_{0})$ depends in general on both $T$ and $\rho_{0}$, 
the Bjerrum length $\lambda_{B}=e^2/\epsilon\kbt$, with $k_{B}$ the Boltzmann constant, could vary through the system. We choose to ignore this dependence henceforth and focus on aqueous electrolytes at fixed $\epsilon$. We ignore convective fluid flow and (implicitly) assume a fixed atmospheric pressure $p$. 
 
An approximate expression for the reversible temperature rise upon electrode charging was proposed in Ref.~\cite{schiffer2006heat} where the adiabatic EDL buildup was described as the isentropic compression of $2\mathcal{N}$ ideal-gas particles, from the complete electrolyte volume $V_{\rm el}=AL$ to two microscopic layers of thickness $\lambda$ and volume $V_{\lambda}=A\lambda$. 
The reversible temperature rise, from the initial low temperature $T_{L}$ to the final high temperature $T_{H}$, is then easily found by evaluating the total differential of the entropy $S(T,\mathcal{V})$, depending here on the volume $\mathcal{V}$ that varies between $V_{\rm el}$ and $V_{\lambda}$, which results in 
\begin{align}\label{eq1:idealgas}
\displaystyle{\ln \frac{T_{H}}{T_{L}}=\frac{2\mathcal{N}k_{B}}{\varrho c_{p}V_{\rm el}}\ln\frac{L}{\lambda}}, 
\end{align}
with the specific heat capacity $c_{x}\equiv T\left(\partial S/\partial T\right)_{x}/\varrho V_{\rm el}$ at a general fixed variable $x$, and $\varrho$ the electrolyte mass density. The authors of Ref.~\cite{schiffer2006heat} inserted EDL characteristics via a Helmholtz model where the number $\mathcal{N}=\mathcal{N}_{H}$ ions involved in the adsorption process scales linearly with the surface charge as $\mathcal{N}_{H}= A\sigma$. Moreover, these ions are confined to a layer of fixed width independent of temperature and salt concentration. A first step towards the inclusion of ion interactions can be made by employing Gouy-Chapman results instead. The number of adsorbed ions is then $\mathcal{N}=\mathcal{N}_{\rm GC}$, with $\mathcal{N}_{\rm GC}/A=\sqrt{\sigma^{2}+\bar{\sigma}^{2}}-\bar{\sigma}$, where $\bar{\sigma}\equiv\sqrt{2\rho_{0}/\pi\lambda_{B}}$, and the EDL width is characterized by the Debye length $\lambda\Rightarrow\lambda_{D}\equiv\sqrt{8\pi \rho_{0}\lambda_{B}}^{-1}$; i.e., $\lambda$ in Eq.~(\ref{eq1:idealgas}) depends on $T$ and $\rho_{0}$ \cite{boon2011blue}. Moreover, $\rho_{0}$ itself depends on the surface charge since the relation $N=\rho_{0}V_{\rm el}+\mathcal{N}$ needs to be obeyed for canonical (fixed $N$) charging.

Alternatively, we now describe the EDLC in terms of the macroscopic variables temperature, charge, and potential from the start. The entropy  $S(T,Q)$ and the potential $\Psi(T,Q)$ are then functions of the independent variables $Q$ and $T$. Since no heat $\delta\mathbb{Q}$ flows through the adiabatic walls of our system, the first law of thermodynamics $dU=\delta\mathbb{Q}+\delta W$ simplifies and the internal energy is solely affected by electrostatic work $\delta W$ performed on the system by the external battery; hence, $dU=\delta W=2\Psi dQ$. The temperature change due to an isentropic change of surface charge now follows from the total differential $dS(T,Q)=0$. Employing a Maxwell relation we find
\begin{align}\label{eq:2adiabat}
d\ln T=\frac{2}{\varrho c_{Q}L}\left(\frac{\partial \Psi}{\partial T}\right)_{Q}ed\sigma.
\end{align}
For aqueous electrolytes at moderate ion concentration the heat capacity of the water molecules dwarfs the heat capacity of the ions. At isobaric conditions this means $c_{Q}\approx c_{p}$, with $c_{p}$ the specific heat capacity of the solvent.

As $\left(\partial \Psi/\partial T\right)_{Q}$ in general depends nontrivially on both $Q$ and $T$, we need to integrate Eq.~(\ref{eq:2adiabat}) numerically, using a relation between the macroscopic observables $\Psi$ and $Q$ for a given electrode and electrolyte system, which can involve experiments \cite{hartel2015heat}, simulations \cite{boda1999low}, or a microscopic model \cite{reszko2005temperature}. In this Letter we capture the EDL within classical density functional theory (DFT). While much effort has been devoted to the development of accurate functionals for the EDL \cite{jiang2014revisiting}, for the illustrative purpose of this Letter it suffices to use a relatively simple grand potential functional $\Omega [\rho_{\pm},\sigma]$, which reads in the planar geometry of interest 
\begin{eqnarray}\label{eq:grandpotential}
\beta\Omega=&A\int_{0}^{L}dz\Big\{\sum\limits_{\alpha=\pm}\rho_{\alpha}(z)\left[\ln \rho_{\alpha} (z)\Lambda_{\alpha}^{3}-1-\beta\tilde{\mu}_{\alpha}(z)\right]\nn
&\hspace{1cm}+\rho_{w}(z)\left[\ln \rho_{w}(z)v-1\right]+\frac{1}{2}\phi(z)q(z)\Big\},
\end{eqnarray}
with $\beta=1/\kbt$ the inverse temperature.
The first line is the ideal-gas grand potential of ions at electrochemical potential $\tilde{\mu}_{\pm}$, with $\Lambda_{\pm}$ the ionic thermal wavelength. The first term of the second line captures steric hindrance qualitatively and is based on a lattice gas model of equal-sized solvent molecules and ions where an upper limit $1/v$ is imposed on the local density via the constraint $[\rho_{+}(z)+\rho_{-}(z)+\rho_{w}(z)]v=1$ \cite{borukhov1997steric}, with $\rho_{w}$ the water density and $v$ the hydrated ionic volume. Finally, the last term in Eq.~(\ref{eq:grandpotential}) is the mean-field electrostatic energy, with $q(z)=\rho_{+}(z)-\rho_{-}(z)+\sigma\left[\delta(z)-\delta(z-L)\right]$ the local unit charge density, and $\phi(z)= e\psi(z)/\kbt$ the local dimensionless electrostatic potential, governed by Poisson's law
\begin{align}\label{eq:Poisson}
\epsilon\partial^{2}_{z}\psi(z)&=-4\pi e\left[\rho_{+}(z)-\rho_{-}(z)\right],\hspace{0.9cm}0<z<L.
\end{align}
The boundary conditions $ \beta e\partial_{z}\psi(z)|_{z=0,L}=-4\pi\lambda_{B} \sigma$ imposed at the electrode surfaces follow from Gauss' law. From the Euler-Lagrange equations $\delta \Omega/\delta \rho_{\pm}(z)=0$ follows 
\begin{align}\label{eq:electrochemicalpotential}
\tilde{\mu}_{\pm}(z)&=\kbt\ln \frac{\rho_{\pm} (z)\Lambda_{\pm}^{3}}{1-v[\rho_{+}(z)+\rho_{-}(z)]}\pm e\psi(z),
\end{align}
which for future reference we split up as $\tilde{\mu}_{\pm}\equiv \mu_{\pm}\pm e\psi$ into the chemical potentials $\mu_{\pm}$, including contributions from all nonelectric interactions, and an electric contribution $e\psi$. Demanding the electrochemical potential $\tilde{\mu}_{\pm}(z)$ to be a spatial constant in equilibrium, we can solve Eq.~(\ref{eq:electrochemicalpotential}) analytically for the density profiles to find modified Boltzmann distributions.  Equation~(\ref{eq:Poisson}) can then be closed yielding the so-called modified Poisson-Boltzmann equation \cite{bikermanphil}, which we solve at a set of temperatures and fixed $\sigma$ to extract the temperature dependence of the surface potential $\Psi=\psi(z=0)$. One then evaluates $\left(\partial \Psi/\partial T\right)_{Q}$ to solve Eq.~(\ref{eq:2adiabat}) for $T_{H}$.

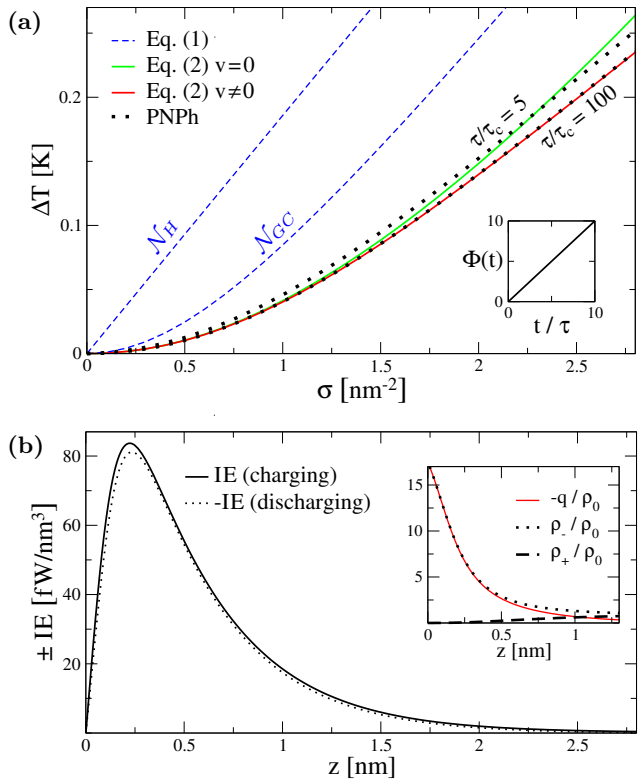
\begin{figure}
\centering
\begin{picture}(8.44,10.5)
\put(0.3,5.0){\includegraphics[width=8.1cm]{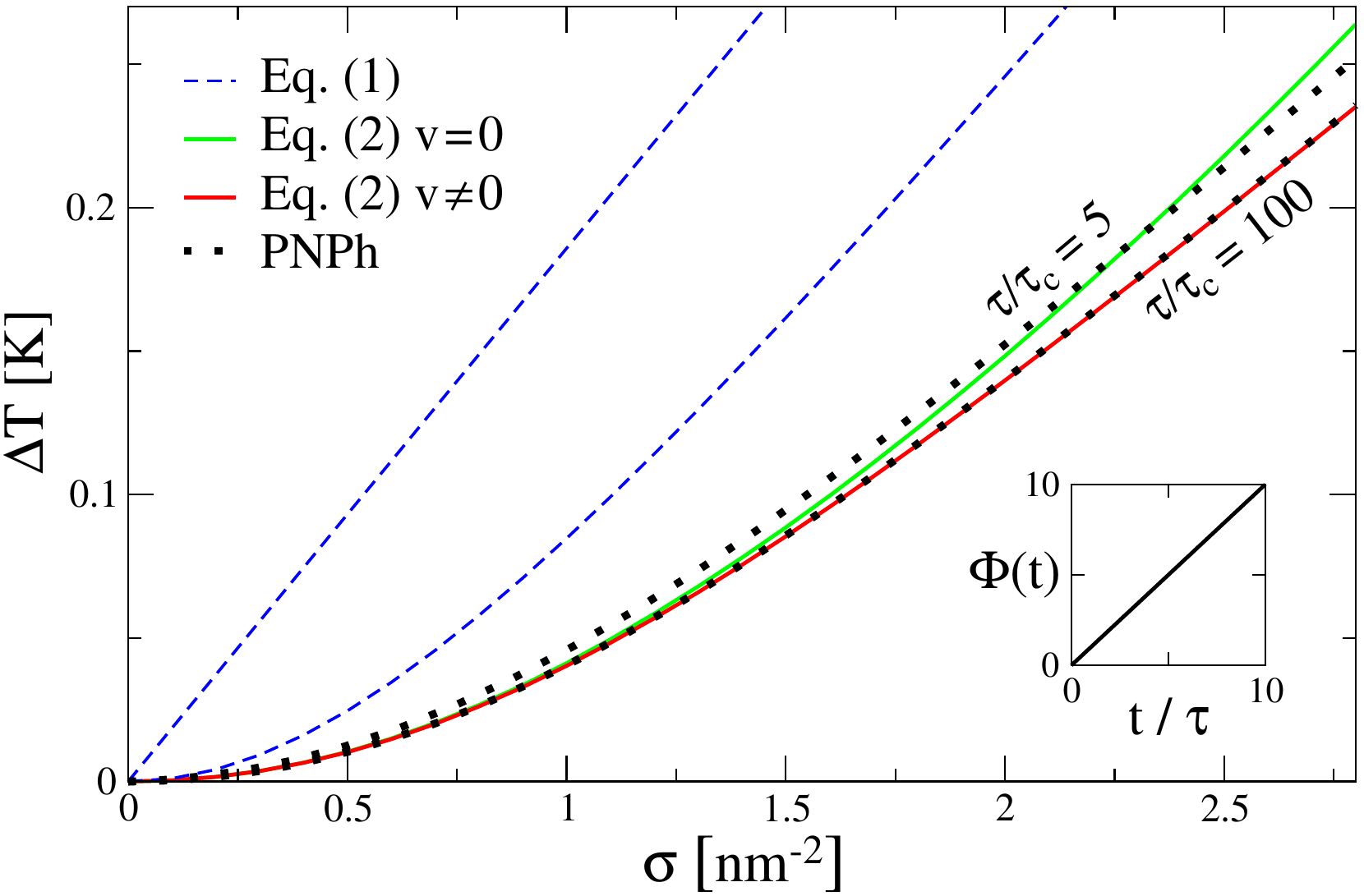}}
\put(0.3,0.0){\includegraphics[width=8.1cm]{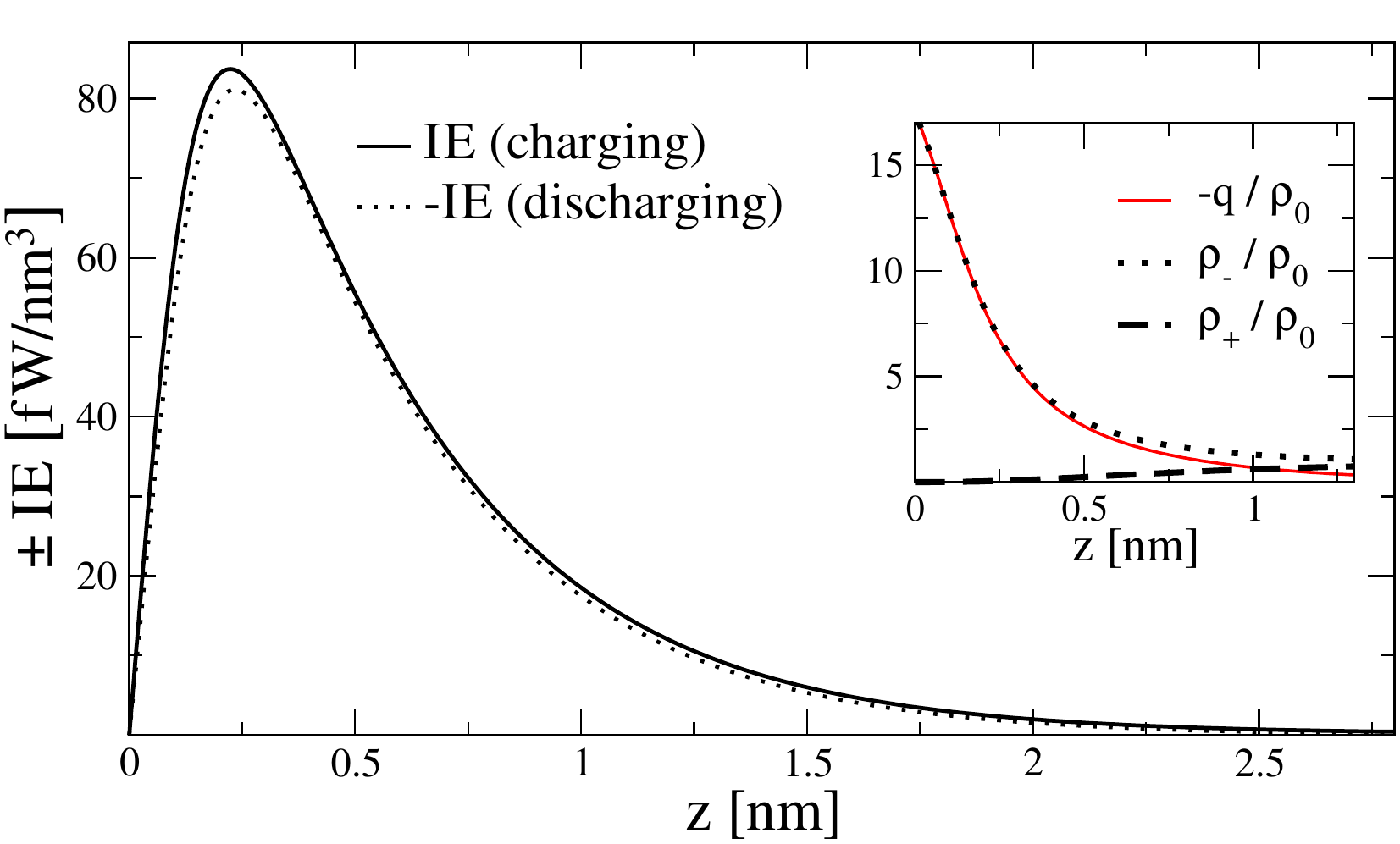}}
\put(0.0,10.0){{\bf (a)}}
\put(0.0,4.35){{\bf (b)}}
\put(2.0,7.0){\begin{rotate}{57} \blue{$\mathcal{N}_{H}$}\end{rotate}}
\put(3.4,7.0){\begin{rotate}{50} \blue{$\mathcal{N}_{GC}$}\end{rotate}}
\end{picture}
\caption{(a) The temperature increase $\Delta T$ upon adiabatic charging of two electrodes separated by $L=50$~nm, starting at uniform salt concentration $\rho_{0}=0.3$ nm$^{-3}$ and a low temperature $T=\SI{20}{\degreeCelsius}$. Plotted are data obtained from Eq.~(\ref{eq1:idealgas}) (blue-dashed lines) with $\lambda=\lambda_{D}$ for both Helmholtz and Gouy-Chapman adsorption, Eq.~(\ref{eq:2adiabat}) for $v=0.16$~nm$^{3}$ (red line) and $v=0$ (green line), as well as the PNPh system (black dotted) with a linear voltage ramp (inset) of inverse slope $\tau/\tau_{c}=\{5, 100\}$.  (b) The total heating rate (line) within the immediate vicinity of the electrode, halfway ($t/\tau=5$) through the $\tau/\tau_{c}=5$ charging process of (a).  The inset shows the corresponding instantaneous anion (line), cation (dashed) and charge density profiles (red dotted). The mirror discharging process at the same time and charging rate is also plotted (dotted). }\label{fig2}
\end{figure}

We turn the discussion to charging dynamics where a time-dependent surface potential $\Psi(t)$ drives the system out of equilibrium. The densities $\rho_{\pm}(z,t)$ and electrostatic potential $\psi(z,t)$ are now time dependent. Moreover, out of equilibrium, the electrochemical potential is not a spatial constant. Consequently, the Poisson equation (still valid since the electromagnetic field responds instantaneously to ``slow" ions) is no longer closed by the Boltzmann weights. We use dynamical DFT  to obtain the ion currents $J_{\pm}=-D\rho_{\pm}\beta\partial_{z} \tilde{\mu}_{\pm}$ from the electrochemical potentials. The diffusion constant $D$ is assumed constant and identical for cations and anions, and for brevity the argument $(z,t)$ is dropped. The ion densities are determined by the continuity equation, to give the Nernst-Planck equation \cite{kilic2007stericII},
$\partial_{t} \rho_{\pm}=D\partial_{z} (\rho_{\pm}\beta\partial_{z} \tilde{\mu}_{\pm})$,
with blocking boundary conditions $J_{\pm}|_{z=0,L}=0$ at the electrodes. The ionic conduction current $I\equiv e\left(J_{+}-J_{-}\right)$ amounts to
\begin{align}\label{eq:ioniccurrent}
I=&-De\Big\{\partial_{z}q+(\rho_{+}+\rho_{-})\beta e\partial_{z}\psi\nn
&\hspace{1cm}+q\partial_{z}\ln \left[1-v(\rho_{+}+\rho_{-}) \right]\Big\}.
\end{align}
Clearly, in the bulk ($q=0$) the electric field $-\partial_{z}\psi$ drives an ionic current $I$ subject to an ionic resistivity $\mathbbl{r}=\kbt/De^{2}(\rho_{+}+\rho_{-})$. Crucial to the reversible heat effect is that the EDL ($\partial_{z}q\neq0$) can support ionic currents that oppose the local electric field.
To find the temperature profiles $T(z,t)$ we need to solve the heat equation (for a derivation see Appendix~\ref{ap2:derivation})
\begin{align}\label{eq:heatequation}
\varrho c_{p}\partial_{t} T&=\kappa\partial_{z}^{2} T+IE.
\end{align}
Here, $\kappa$ is the heat conductivity, and the source term $IE\equiv\mathbbl{\dot{q}}_{\rm irr}+\mathbbl{\dot{q}}_{\rm rev}$
consists of the (ir)reversible heating rates $\mathbbl{\dot{q}}_{\rm rev}= I\mathbbl{r} De\left\{\partial_{z}q+q\partial_{z}\ln \left[1-v(\rho_{+}+\rho_{-}) \right]\right\}$ and $\mathbbl{\dot{q}}_{\rm irr}= I^2 \mathbbl{r}$. 
Note that $\mathbbl{\dot{q}}_{\rm rev}$ is nonvanishing only in the EDL, where $q\neq0$. Moreover, within this region the ratio $\mathbbl{\dot{q}}_{\rm irr}/\mathbbl{\dot{q}}_{\rm rev}\sim I/\partial_{z}q\to0$ for slow charging ($I\to0$).  
Equations (\ref{eq:Poisson})-(\ref{eq:ioniccurrent}) and (\ref{eq:heatequation}) form the closed set PNPh, for Poisson, Nernst-Planck, and heat. Numerical results for the $(z,t)$ dependence of $\phi,\rho_{\pm}, q, J_{\pm},I$, and $T$ were obtained for an (initially uncharged) EDLC of plate separation $L=50$~nm at $T=\SI{20}{\degreeCelsius}$, filled with an aqueous NaCl electrolyte at $\rho_{0}=0.3$ nm$^{-3}$. We use the following parameter set $v=0.16$~nm$^{3}$ (from Ref.~\cite{nightingale1959phenomenological}), $D=1.6\times10^{-9}$~m$^{2}$ s$^{-1}$, $\kappa=0.599$~W m$^{-1}$ K$^{-1}$, $\epsilon=71$, $\varrho=998.3$ kg m$^{-3}$, $c_{p}=4.182$~kJ K$^{-1}$ kg$^{-1}$. We start with the uncharged EDLC at $T_{L}=\SI{20}{\degreeCelsius}$ and ramp the dimensionless electrode potential $\Phi\equiv\phi(z=0)$ linearly from $\Phi=0$ to $\Phi=10$ during $10 \tau$ [Fig.~\ref{fig2}(a), inset]. For the slow charging rates $\tau/\tau_{c}=\{5, 100\}$ considered, with time measured in units of the ``RC time'' $\tau_{c}=\lambda_{D}L/D$, the temperature $T(z,t)$ is practically homogenous throughout the cell. With $\sigma(t)$ and $T(z,t)$ at hand we eliminate the time dependence of the latter. The black dotted lines in Fig.~\ref{fig2}(a) represent the measurement $\Delta T(\sigma)=T(z=0,\sigma)-T_{L}$.
We also plot the adiabatic temperature rise as predicted by the thermodynamic identities Eq.~(\ref{eq1:idealgas}) for both $\mathcal{N}_{GC}$ and $\mathcal{N}_{H}$ (blue-dashed lines), and Eq.~(\ref{eq:2adiabat}) 
(red line). For the slow charging process at $\tau/\tau_{c}=100$ we see a near-perfect agreement between the temperature rise predictions of Eq.~(\ref{eq:2adiabat}) and the PNPh equations. Equation~(\ref{eq1:idealgas}) does not perform as well. This is numerical evidence for our claim that Eq.~(\ref{eq:2adiabat}) and {\it not} Eq.~(\ref{eq1:idealgas}) captures the thermodynamics of EDLCs. 
We ascribe the small temperature rise at  $\sigma\ll\bar{\sigma}$ (here, $\bar{\sigma}=0.49$ nm$^{-2}$) to ``ion swapping": at low potentials, surface charge is screened via both coion repulsion and counterion attraction, yielding inefficient salt adsorption.  Conversely, at higher potentials, coions are depleted from the electrodes' vicinity, such that each additional electron (hole) attracts an additional counterion. Since the adiabatic temperature rise is driven by the surface charge-induced electrolyte inhomogeneity, the nonlinear screening regime $\sigma \gg \bar{\sigma}$ yields a higher differential temperature increase.
Accordingly, $\mathcal{N}_{GC}$  constitutes a considerable improvement over $\mathcal{N}_{H}$ for Eq.~(\ref{eq1:idealgas}), since it incorporates this transition from $\mathcal{N}_{GC}\sim \sigma^2/2\bar{\sigma}$ at $\sigma\ll\bar{\sigma}$, to  $\mathcal{N}_{GC}\sim \sigma$ at $\sigma\gg\bar{\sigma}$. 
For comparison, we also evaluated Eq.~(\ref{eq:2adiabat}) for regular ($v=0$) Poisson-Boltzmann theory (green line). The influence of ion size only shows at higher potentials, when packing constraints start to affect the electrochemical potential Eq.~(\ref{eq:electrochemicalpotential}). The higher temperature variations predicted at $v=0$ indicate that incorporation of ionic volume lowers the entropic contribution to the grand potential \cite{kralj1996simple}.

In Fig.~\ref{fig2}(b) we show the total heating rate $IE$ halfway ($t/\tau=5$) through the charging process with $\tau/\tau_{c}=5$. The corresponding instantaneous anion (dotted), cation (dashed) and charge density profiles (red) are shown in Fig.~\ref{fig2}(b) (inset). For comparison we also consider the reversed process, starting at an equilibrated state at $\Phi=10$ and discharging to $\Phi=0$ during $10\tau$ with $\tau/\tau_{c}=5$. The negative of the heating rate halfway through this discharging process is indicated with a dotted line in Fig.~\ref{fig2}(b). The heating rates $IE$ exhibit a clear peak associated with $\mathbbl{\dot{q}}_{\rm rev}$ in the EDL where $q$ is nonvanishing. Towards the bulk only the small contribution $\mathbbl{\dot{q}}_{\rm irr}$ persists. Upon decreasing the (dis)charging rates, this strictly positive Joule heating gets progressively smaller, and the (dis)charging heating rates turn into mirror images of one another (not shown).

This Letter discusses reversible heating in EDLCs from two different viewpoints. On the one hand we considered the heat equation, which was derived decades ago for general settings \cite{groot1962non}. Only recently \cite{d2014first} it was specified to the case of adiabatic EDLC charging, with ion currents captured by Poisson-Nernst-Planck equations. This model was shown to quantitatively reproduce the reversible temperature oscillation as observed in supercapacitors \cite{schiffer2006heat}. The PNPh model exhibits cooling where ionic currents, adhering to the gradient in electrochemical potential, {\it oppose} the local electric field. 
Since this only occurs in the EDL, the reversible heating effect is highly localized [see Fig.~\ref{fig2}(b)]. 
While the level of sophistication of the PNPh model sufficed for the purposes of this Letter (large plate separation, slow charging, hence, small spatial temperature variation), when considering fast (dis)charging of nanoporous supercapacitors the adiabatic approximation underlying dynamical DFT becomes less justifiable. Moreover, when spatial variations in the diffusion constant and temperature become nonnegligible, the use of a free energy functional is problematic, as in DFT the temperature enters as an  imposed (spatially constant) parameter. Future work could build on recent developments that address these problems \cite{schmidt2013power, schmidt2011statics, kondrat2015dynamics}.

The other, thermodynamic, viewpoint, brought fourth two distinct identities [Eqs.~(\ref{eq1:idealgas}) and (\ref{eq:2adiabat})] for the temperature change upon isentropic charging of EDLCs. 
Compared to the PNPh model, the merits of Eq.~(\ref{eq:2adiabat}) are twofold. Firstly, its simplicity aides interpretation. Reversible temperature changes are controlled essentially by a small set of parameters $\{ \sigma, \varrho c_{Q}, L\}$ together with a system-dependent derivative $\left(\partial \Psi/\partial T\right)_{Q}$. 
The second merit of Eq.~(\ref{eq:2adiabat}) is that, as a thermodynamic identity, it does not rely on uncontrolled approximations, and can be used as a reliable predictor for the lower bound of temperature variations. Approximations enter the theory at the level of $\left(\partial \Psi/\partial T\right)_{Q}$, so that more accurate estimates can be found by systematically improving the grand potential Eq.~(\ref{eq:grandpotential}), by including, for instance, solvent polarizability \cite{hatlo2012electric}, a better description of excluded volume interactions \cite{hartel2015fundamental}, and residual ion correlations \cite{jiang2014revisiting} (see also Appendix~\ref{ap3:correlations}).  

Though oversimplified ideal-gas reasoning permeates the reversible heating literature \cite{xiong2015thermal, zhang2016reversible, gualous2011supercapacitor, d2014first, lee2014modeling}, the main finding of this Letter is that the thermodynamic identity Eq.~(\ref{eq:2adiabat}), and {\it not} Eq.~(\ref{eq1:idealgas}), constitutes the slow charging limit of the PNPh system. The kinetic and thermodynamic viewpoints give complementary information that together allow for a thorough understanding of reversible temperature variations in EDLCs.  
While structural transitions of the EDL in supercapacitors under \textit{isothermal} conditions are the subject of intense study  \cite{chmiola2006anomalous}, the findings of this Letter should form the basis for understanding  heat effects that such structural rearrangements undoubtedly induce in adiabatic setups. 
Adiabatic temperature measurements as discussed here probe thermodynamic response 
that is not isothermally accessible. The thermal response of adiabatically (dis)charged supercapacitors therefore carries  information that could deepen our understanding of the electric double layer.


\begin{acknowledgments}
This work is part of the D-ITP consortium, a program of the 
Netherlands Organisation for Scientific Research (NWO) that is funded by 
the Dutch Ministry of Education, Culture and Science (OCW). 
We acknowledge financial support from an NWO-VICI grant, 
and thank Bram Bet, Ben Ern\'{e}, Andreas H\"{a}rtel, Yan Levin, Jos\'{e} A. Manzanares, and
Paul van der Schoot for useful discussions. 
\end{acknowledgments}

\begin{appendix}

\section{Derivation heat equation}\label{ap2:derivation}
To find the heat equation Eq.~(\ref{eq:heatequation}), 
we follow Ref.~\cite{groot1962non} and define the local specific internal energy $u\equiv U/\varrho V_{\rm el}$ as the energy associated with thermal agitation and all short ranged (nonelectrostatic) particle interactions. We can now write the internal energy balance as 
\begin{eqnarray}\label{eq:energy_balance}
\varrho \partial_{t} u&=&-\partial_{z} J_{q}+IE,
\end{eqnarray}
or, similarly, the internal enthalpy balance,
\begin{equation}\label{eq:enthalpy_balance2}
\varrho\partial_{t}  h=-\partial_{z}  J_{q}+\partial_{t} p+IE,
\end{equation}
where the local specific enthalpy $h$ is defined via $\varrho h=\varrho u+p$.
In these equations, the heat flow $J_{q}=-\kappa \partial_{z} T+\sum_{i}\bar{H}_{i} J_{i}$ contains Fourier heat diffusion, and the partial molecular enthalpy $\bar{H}_{i}$ carried by particle currents.
Note that we did not include a Dufour term $\sim\nabla \mu$ in $J_{q}$, which is consistent with our disregarding Soret terms $\sim\nabla T$, the reciprocal phenomenon, in the Nernst-Planck equation 
\cite{curtiss1999multicomponent}.

\subsection{ Partial molecular enthalpy}
The partial molecular enthalpy $\bar{H}_{i}$ of component $i\in\{+,-,s\}$ is defined for homogeneous systems in terms of the enthalpy $H(S,p,N_{\pm},N_{s})$  as
\begin{align}\label{eq:partialmolecularenthalpydefinition}
\bar{H}_{i}&\equiv \left(\frac{\partial H}{\partial N_{i}}\right)_{T,p,N_{i'\neq i}},
\end{align}
which is related to the partial {\it molar} enthalpy (common in chemistry literature) by division by Avogadro's number.
Above, we defined the internal energy as the kinetic energy and microscopic interaction energy of the constituent particles {\it without} electric contributions, which is in line with Refs.~\cite{groot1962non,haase1968thermodynamics,kontturi2008ionic}.
With this choice, Refs.~\cite{de1953thermodynamics, groot1962non, kontturi2008ionic}  argue that it is the chemical potential, and not the electrochemical potential that enters the Gibbs relation. Therefore, the total differential of the enthalpy reads
\begin{align}\label{eq:differentialenthalpy}
dH&=TdS+Vdp+\sum_{i}\mu_{i}dN_{i}.
\end{align}
Using the total differential of the entropy (employing a Maxwell identity and identifying the heat capacity $C_{p}$),
\begin{align}\label{eq:differentialentropy}
dS
&=\frac{C_{p}}{T}dT-\left(\frac{\partial V}{\partial T}\right)_{p,N_{i}}dp-\sum_{i}\left(\frac{\partial \mu_{i}}{\partial T}\right)_{p,N_{i}}dN_{i},
\end{align}
we eliminate $dS$ in Eq.~(\ref{eq:differentialenthalpy}) in favor of $dT$ to find
\begin{align}\label{eq:differentialenthalpyT}
dH=&C_{p} dT+\left[V-T\left(\frac{\partial V}{\partial  T}\right)_{p,N_{i}} \right]dp\nn
&+\sum_{i}\left[\mu_{i}-T\left(\frac{\partial \mu_{i}}{\partial T}\right)_{p,N_{i}}\right]dN_{i}.
\end{align}
The partial molecular enthalpy Eq.~(\ref{eq:partialmolecularenthalpydefinition}) then reduces to the so-called partial Gibbs-Helmholtz equation 
\begin{align}\label{eq:derivationgibsshelmholtz}
\bar{H}_{i}&=\mu_{i}-T\left(\frac{\partial \mu_{i}}{\partial T}\right)_{p,N_{i'}}=-T^{2}\left(\frac{\partial \mu_{i}/T}{\partial T}\right)_{p,N_{i'}},
\end{align}
with $i'\in\{+,-,s\}$. 
Since our system of interest is {\it not} homogenous, one should instead consider a subspace region of volume $\mathcal{V}$, carrying entropy $\mathcal{S}$ and occupied by $\mathcal{N}_{i}$ particles of each of the $i$ species. This region should be small enough so that $\rho_{i}=\mathcal{N}_{i}/\mathcal{V}$ is the locally homogenous particle density. One could then repeat the above exercise to find the partial molecular enthalpy  $\bar{H}_{i}\equiv \left(\partial \mathcal{H}/\partial \mathcal{N}_{i}\right)_{T,p,\mathcal{N}_{i'\neq i}}$, defined in terms of the enthalpy $\mathcal{H}(\mathcal{S},p, \mathcal{N}_{\pm}, \mathcal{N}_{s})$ within this space region, to find the same expression Eq.~(\ref{eq:derivationgibsshelmholtz}) in terms of the $z$-dependent chemical potential.
Inserting the ionic chemical potential (first part of Eq.~(\ref{eq:electrochemicalpotential})) into Eq.~(\ref{eq:derivationgibsshelmholtz}) then gives
\begin{align}\label{eq:partialmolarenthalpycontinued}
\bar{H}_{\pm}
=&\frac{3}{2}\kbt+\frac{\kbt}{1-v(\rho_{+}+\rho_{-})}\left(\frac{\partial \ln \mathcal{V}}{\partial \ln T} \right)_{p}.
\end{align}
were we used that $\Lambda_{i}\sim 1/\sqrt{T}$. Interestingly, the first term in Eq.~(\ref{eq:partialmolarenthalpycontinued}) that appears is the ideal-gas energy, and {\it not} the ideal-gas enthalpy.
If all species $i$ had been treated at the ideal-gas-level (with $v=0$), then one should have substituted $\rho_{i}=p/\kbt$ in Eq.~(\ref{eq:electrochemicalpotential}), which would have led to the ideal-gas enthalpy $\bar{H}_{i}=5\kbt/2$. 
However, adding ions (see Eq.~(\ref{eq:partialmolecularenthalpydefinition})), the volume they explore does not grow as an ideal gas at fixed pressure as $\mathcal{V}\sim \mathcal{N}_{\pm}$, instead $\mathcal{V}$ stays roughly unaffected because it  is primarily determined by the solvent molecules, which are markedly nonideal. 

In the second term of Eq.~(\ref{eq:partialmolarenthalpycontinued}) we recognize $\left(\partial \ln \mathcal{V}/\partial \ln T \right)_{p}=\alpha T$ with $\alpha$ the volumetric expansivity $\alpha\equiv  \left(\partial \ln \varrho/\partial T \right)_{p}$ of the fluid at mass density $\varrho$.
In the bulk electrolyte, the volume $\mathcal{V}$ is predominantly occupied by water molecules, so that we can interpret the volumetric expansivity of the fluid with that of pure water. At $\SI{20}{\degreeCelsius}$ this amounts to $\alpha T=0.06$ \cite{kell1975density}. The second term in Eq.~(\ref{eq:partialmolarenthalpycontinued}) therefore only gives a small correction to the first term, and is often disregarded \cite{davidson1962}. 
As opposed to the dilute bulk, at high electric potentials  
Eq.~(\ref{eq:partialmolarenthalpycontinued}) is problematic in the EDL. Interpreting the coefficient of thermal expansivity $\alpha$ with that of pure water does not hold here because of high local ion densities. Moreover, an artifact of the lattice-gas free energy functional Eq.~(\ref{eq:grandpotential}), 
the steric interactions are a-thermal (do not depend on temperature). This is also reflected by the {\it fixed} lattice spacing $v^{1/3}$ which erroneously always gives $(\partial \ln \mathcal{V}/\partial \ln T)_{p}=0$. Meanwhile, the vanishing of the term $(\partial \ln \mathcal{V}/\partial \ln T)_{p}$ is at high electric potentials accompanied by a divergence of its prefactor $\kbt/[1-v(\rho_{+}+\rho_{-})]$. More explicit solvent modeling might be necessary to get a better grip on the thermal expansivity term of $\bar{H}_{i}$ in the EDL region.  

\subsection{Internal enthalpy balance}
With the partial molecular enthalpy at hand, we now set out to derive the heat equation, which we do by finding an alternative expression for the l.h.s. of the internal enthalpy balance Eq.~(\ref{eq:enthalpy_balance2}).
Similar treatments can be found in Refs. \cite{haase1968thermodynamics,groot1962non,bird2007transport}. 

We write the total mass of component $i$ as  $m_{i}=N_{i}M_{i}$, with $M_{i}$ the molecular weight, such that the total mass is $m=m_{+}+m_{-}+m_{s}$ and the electrolyte mass density is $\varrho=m/V_{\rm el}$. The enthalpy $H(S,p,N_{\pm},N_{s})$ can then be written in terms of the total mass of the individual components $H(m_{+},m_{-},m_{s})$. Euler's theorem then allows us to write the enthalpy in terms mass fractions $c_{\pm}=m_{\pm}/m$ as $H(m_{+},m_{-},m_{s})=m h(c_{+},c_{-})$, with  $h$ the specific enthalpy density, the total differential of which reads
\begin{align}\label{eq:specificenthalpydifferential}
\varrho \frac{dh}{dt}=&\varrho\left(\frac{\partial h}{\partial T}\right)_{p,c_{k}}\frac{dT}{dt}+\varrho\left(\frac{\partial h}{\partial p}\right)_{T,c_{k}}\frac{dp}{dt}\nn
&+\sum_{k}\varrho\left(\frac{\partial h}{\partial c_{k}}\right)_{T,p,c_{k'\neq k}}\frac{dc_{k}}{dt},
\end{align}
with $k\in\{+,-\}$.
For the last term Ref.~\cite{groot1962non} (p. 458) and Ref.~\cite{bird2007transport} (p. 609) provide alternative derivations both yielding 
\begin{align}
\left(\frac{\partial h}{\partial c_{k}}\right)_{T,p}=\frac{ \bar{H}_{k}}{M_{k}}-\frac{\bar{H}_{s}}{M_{s}}.
\end{align}
The continuity equation for mass fluxes reads  $\varrho(dc_{i}/dt)=-\partial_{z} j_{i}$, in terms of the mass flux $j_{i}$, which is related to the particle flux as $j_{i}=M_{i}J_{i}$. The absence of barycentric motion implies that we can replace the material derivatives with partial derivatives, and also that the mass fluxes obey $\sum_{i}j_{i}=0$. The above considerations  yield
\begin{align}
\sum_{k}\varrho\left(\frac{\partial h}{\partial c_{k}}\right)_{T,p,c_{k'\neq k}}\frac{dc_{k}}{dt}=\sum_{i}\bar{H}_{i}\partial_{z} j_{i}.
\end{align}
 The first two partial derivatives in Eq.~(\ref{eq:specificenthalpydifferential}) can be easily identified in Eq.~(\ref{eq:differentialenthalpyT}) to find
\begin{align}\label{eq:enthalpy_balance1}
\varrho\partial_{t}h=&\varrho c_{p}\partial_{t}T+\left[1-\left(\frac{\partial \ln V}{\partial \ln T}\right)_{p} \right]\partial_{t}p-\sum_{i} \bar{H}_{i}\partial_{z} J_{i},
\end{align}
which is the alternative expression for the l.h.s. of the internal enthalpy balance Eq.~(\ref{eq:enthalpy_balance2}) we referred to at the start of this subsection.

\subsection{Heat equation}
Combination of Eqs.~(\ref{eq:enthalpy_balance2}) and (\ref{eq:enthalpy_balance1}), and inserting the heat flow, $J_{q}=-\kappa \partial_{z} T+\sum_{i}\bar{H}_{i} J_{i}$, then gives 
\begin{align}\label{eq:fullheatequation}
\varrho c_{p}\partial_{t}T=\kappa\partial_{z}^{2} T+IE+\alpha T \partial_{t}p-\sum_{i} J_{i}\partial_{z}\bar{H}_{i}. 
\end{align}
Here, the electric field is found via Eq.~(\ref{eq:ioniccurrent}) as
\begin{align}\label{eq:electricfield}
E=&I\mathbbl{r}+\mathbbl{r}De\left(\partial_{z}q+q\partial_{z}\ln \left[1-v(\rho_{+}+\rho_{-}) \right]\right).
\end{align}
with $\mathbbl{r}=\kbt/(De^{2}(\rho_{+}+\rho_{-}))$. The term $IE$ then gives the (ir)reversible heating rates $\mathbbl{\dot{q}}_{\rm irr}\equiv I^2 \mathbbl{r}$ and $\mathbbl{\dot{q}}_{\rm rev}\equiv I\mathbbl{r} De\left\{\partial_{z}q+q\partial_{z}\ln \left[1-v(\rho_{+}+\rho_{-}) \right]\right\}$.

As discussed above, the term $\alpha\equiv  \left(\partial \ln \varrho/\partial T \right)_{p}$ in Eq.~(\ref{eq:fullheatequation}) is the (usually small) volumetric thermal expansivity of the electrolyte. Also $\partial_{t}p$ is small since our (incompressible) liquid is isobaric throughout the cell, and from here on we therefore drop the term $\alpha T \partial_{t} p$.  

For the ionic contribution to the term $J_{i}\nabla \bar{H}_{i} $ consider the gradient of  Eq.~(\ref{eq:partialmolarenthalpycontinued}). 
Firstly, the term $\sim \partial_{z}T$ vanishes in equilibrium. 
Meanwhile, Eq.~(\ref{eq:derivationgibsshelmholtz}) also contains a term proportional to the thermal expansivity of the solvent $\sim \partial_{T} \ln \mathcal{V}$, the gradient of which vanishes in the bulk. In equilibrium $\partial_{z}\bar{H}_{s}=0$ because the solvent chemical potential is uniform though the cell. All these terms vanishing in equilibrium implies that $J_{i}\partial_{z} H_{i}$ is proportional to $\sim J_{i}^{2}$ (or higher powers in $J_{i}$) and therefore vanishes faster than reversible contributions that go as $\mathbbl{\dot{q}}_{\rm rev}\sim I$.  We conclude that  $J_{i}\partial_{z} H_{i}$ does {\it not} contribute to reversible heating. Moreover, we see that the ratio $(J_{+}+J_{-})\nabla \kbt/\mathbbl{\dot{q}}_{\rm irr}$ also goes to zero since for the system of interest both the neutral salt current $(J_{+}+J_{-})$ and temperature variations $\nabla T$ are very small. We therefore omit $J_{i}\nabla \bar{H}_{i} $ from Eq.~(\ref{eq:fullheatequation}) that now simplifies to the heat equation Eq.~(\ref{eq:heatequation}) 
\begin{align}
\varrho c_{p}\partial_{t}T=\kappa\partial_{z}^{2} T+\mathbbl{\dot{q}}_{\rm irr}+\mathbbl{\dot{q}}_{\rm rev}.
\end{align}
Note that our derivation of this equation differs from the one presented in Ref.~\cite{d2014first}. These authors started from internal energy balance (similar to our Eq.~(\ref{eq:energy_balance}) but) lacking the term $IE$, with this term appearing in the heat equation via the partial molecular enthalpy that they claim to be $\bar{H}_{i}=\tilde{\mu}_{i}-T\left(\partial \mu_{i}/\partial T\right)$.\\ 

\section{Appendix: Analytical approximation to the adiabatic temperature rise}\label{ap4:largeandsmallcharge}
For the case of vanishing ionic volume ($v=0$) and no double-layer overlap, Gouy and Chapman famously found an analytic solution to the Poisson-Boltzmann equation. The electrodes of our model EDLC are sufficiently separated that the ionic density profiles can be considered to be nonoverlapping. Consequently, we can approximate the adiabatic temperature rise predicted within Poisson-Boltzmann theory (the green line in Fig.~\ref{fig2}(a)) by inserting the Gouy-Chapman potential $\Psi= (2\kbt/e) \sinh^{-1}(\sigma/\bar{\sigma})$ into Eq.~(\ref{eq:2adiabat}), which gives
\begin{align}\label{eq:c1}
\frac{1}{T}dT
 &= \frac{4k_{\rm B}}{\varrho c_{Q}L}\left[\sinh^{-1} \frac{\sigma'}{\bar{\sigma}}-\frac{\sigma}{\sqrt{\bar{\sigma}^{2}+\sigma^{2}}}\frac{\partial \ln\bar{\sigma}}{\partial \ln T}\right]d\sigma.
\end{align}
Separation of variables in this equation is only possible for the special case where $\partial_{T}\bar{\sigma}(T)=0$, which occurs if $T\cdot \epsilon(T)=cst$ (i.e., not the case considered in this Letter).  In that case we find
\begin{align}\label{eq4}
\ln\frac{T_H}{T_L}&=\frac{4k_{\rm B}}{\varrho c_{Q}L}\bigg[\sigma\sinh^{-1} \frac{\sigma}{\bar{\sigma}}-\sqrt{\bar{\sigma}^{2}+\sigma^{2}}+\bar{\sigma}\bigg],
\end{align}
with $T_L$ the low initial temperature and $T_H$ the higher temperature after adiabatic charging. For small temperature changes, $\Delta T=T_H-T_L$, this equation simplifies to
\begin{align}\label{eq5}
\Delta T&\approx\frac{4k_{\rm B}T_L}{\varrho c_{Q}L}\times\begin{cases}
\frac{\sigma^2}{2\bar{\sigma}}\quad\quad&\textrm{if~} \,  \sigma\ll\bar{\sigma}, \\
\sigma\ln \frac{2\sigma}{\bar{\sigma}}-\sigma \quad\quad&\textrm{if~} \,  \sigma\gg\bar{\sigma}.
\end{cases}  
\end{align}
While this special case of a $T$-independent $\bar{\sigma}$ reveals the small- and large-$\sigma$ scaling behavior, neglecting the second term in Eq.~(\ref{eq:c1}) leads to an overestimation of the adiabatic temperature rise predicted by Eq.~(\ref{eq4}) by about 40\% w.r.t. the prediction of Eq.~(\ref{eq:c1}) for the parameters chosen.

\section{ Excess correlations}\label{ap3:correlations}
Electrolytic partial molar enthalpy Eq.~(\ref{eq:derivationgibsshelmholtz}) is well-documented in the chemistry literature \cite{davidson1962, devoe2001thermodynamics} for the electro-neutral bulk. Instead of the term proportional to the thermal expansivity of the solvent, discussed at length above, usually excess chemical potential due to ionic correlations are considered, via Debye-H\"{u}ckel (DH) theory where $\beta\mu^{exc}_{\rm DH}(T,\rho_{0})=-\lambda_{B}/2\lambda_{D}$, or extensions thereof that include for instance finite ion size. In the bulk, adding $\mu^{exc}_{\rm DH}$ will not contribute to the reversible heating since both $\partial_{z}T$ and $\partial_{z} \rho_{0}$ vanish in equilibrium.
Meanwhile, the assumptions underlying DH theory are strongly violated in the EDL where same-sign counterions are at high density. We therefore expect the DH heating rates of Ref.~\cite{d2014first}, with a significant nonzero contribution in the EDL (see Fig.~(6) of  Ref.~\cite{d2014first}), to be unreliable. This is substantiated by our Fig.~\ref{fig2}(a) which shows that the solutions to the PNPh model as formulated in this Letter for slow charging coincide with the adiabatic temperature rise as predicted by the thermodynamic identity Eq.~(\ref{eq:2adiabat}). 
Had we added ionic correlations via the formulation of Ref.~\cite{d2014first}, that is, only in the partial molecular enthalpy to only affect the PNPh equations, we would have found only the black dotted lines of Fig.~\ref{fig2}(a) shifted, but not red line of the thermodynamic results of Eq.~(\ref{eq:2adiabat}). If first-principles modeling is desired, excess ion correlations should be incorporated at the level of the grand potential functional Eq.~(\ref{eq:grandpotential}), to impact (via the electrochemical potential) both the EDL and the bulk, in as well as out of equilibrium.

\end{appendix}

\begin{thebibliography}{49}%
\makeatletter
\providecommand \@ifxundefined [1]{%
 \@ifx{#1\undefined}
}%
\providecommand \@ifnum [1]{%
 \ifnum #1\expandafter \@firstoftwo
 \else \expandafter \@secondoftwo
 \fi
}%
\providecommand \@ifx [1]{%
 \ifx #1\expandafter \@firstoftwo
 \else \expandafter \@secondoftwo
 \fi
}%
\providecommand \natexlab [1]{#1}%
\providecommand \enquote  [1]{``#1''}%
\providecommand \bibnamefont  [1]{#1}%
\providecommand \bibfnamefont [1]{#1}%
\providecommand \citenamefont [1]{#1}%
\providecommand \href@noop [0]{\@secondoftwo}%
\providecommand \href [0]{\begingroup \@sanitize@url \@href}%
\providecommand \@href[1]{\@@startlink{#1}\@@href}%
\providecommand \@@href[1]{\endgroup#1\@@endlink}%
\providecommand \@sanitize@url [0]{\catcode `\\12\catcode `\$12\catcode
  `\&12\catcode `\#12\catcode `\^12\catcode `\_12\catcode `\%12\relax}%
\providecommand \@@startlink[1]{}%
\providecommand \@@endlink[0]{}%
\providecommand \url  [0]{\begingroup\@sanitize@url \@url }%
\providecommand \@url [1]{\endgroup\@href {#1}{\urlprefix }}%
\providecommand \urlprefix  [0]{URL }%
\providecommand \Eprint [0]{\href }%
\providecommand \doibase [0]{http://dx.doi.org/}%
\providecommand \selectlanguage [0]{\@gobble}%
\providecommand \bibinfo  [0]{\@secondoftwo}%
\providecommand \bibfield  [0]{\@secondoftwo}%
\providecommand \translation [1]{[#1]}%
\providecommand \BibitemOpen [0]{}%
\providecommand \bibitemStop [0]{}%
\providecommand \bibitemNoStop [0]{.\EOS\space}%
\providecommand \EOS [0]{\spacefactor3000\relax}%
\providecommand \BibitemShut  [1]{\csname bibitem#1\endcsname}%
\let\auto@bib@innerbib\@empty
\bibitem [{\citenamefont {Gouy}(1910)}]{gouy1910constitution}%
  \BibitemOpen
  \bibfield  {author} {\bibinfo {author} {\bibfnamefont {G.}~\bibnamefont
  {Gouy}},\ }\href@noop {} {\bibfield  {journal} {\bibinfo  {journal} {J.
  Phys. Theor. Appl.}\ }\textbf {\bibinfo {volume} {9}},\ \bibinfo {pages} {457} (\bibinfo
  {year} {1910})};
\bibfield  {author} {\bibinfo {author} {\bibfnamefont {D.~L.}~\bibnamefont
  {Chapman}},\ }\href@noop {} {\bibfield  {journal} {\bibinfo  {journal}
  {Philos. Mag.}\ }\textbf {\bibinfo {volume} {25}} (\bibinfo {year}
  {1913})}  
  \BibitemShut {NoStop}%
\bibitem [{\citenamefont {Schiffer}\ \emph {et~al.}(2006)\citenamefont
  {Schiffer}, \citenamefont {Linzen},\ and\ \citenamefont
  {Sauer}}]{schiffer2006heat}%
  \BibitemOpen
  \bibfield  {author} {\bibinfo {author} {\bibfnamefont {J.}~\bibnamefont
  {Schiffer}}, \bibinfo {author} {\bibfnamefont {D.}~\bibnamefont {Linzen}},\
  and\ \bibinfo {author} {\bibfnamefont {D.~U.}\ \bibnamefont {Sauer}},\
  }\href@noop {} {\bibfield  {journal} {\bibinfo  {journal} {J. Power
  Sources}\ }\textbf {\bibinfo {volume} {160}},\ \bibinfo {pages} {765}
  (\bibinfo {year} {2006})}\BibitemShut {NoStop}%
\bibitem [{\citenamefont {Simon}\ and\ \citenamefont
  {Gogotsi}(2008)}]{simon2008materials}%
  \BibitemOpen
  \bibfield  {author} {\bibinfo {author} {\bibfnamefont {P.}~\bibnamefont
  {Simon}}\ and\ \bibinfo {author} {\bibfnamefont {Y.}~\bibnamefont
  {Gogotsi}},\ }\href@noop {} {\bibfield  {journal} {\bibinfo  {journal}
  {Nat. Mater.}\ }\textbf {\bibinfo {volume} {7}},\ \bibinfo {pages} {845}
  (\bibinfo {year} {2008})};
  \bibfield  {author} {\bibinfo {author} {\bibfnamefont {M.}~\bibnamefont
  {Salanne}}, \bibinfo {author} {\bibfnamefont {B.}~\bibnamefont {Rotenberg}},
  \bibinfo {author} {\bibfnamefont {K.}~\bibnamefont {Naoi}}, \bibinfo {author}
  {\bibfnamefont {K.}~\bibnamefont {Kaneko}}, \bibinfo {author} {\bibfnamefont
  {P.-L.}\ \bibnamefont {Taberna}}, \bibinfo {author} {\bibfnamefont
  {C.~P.}~\bibnamefont {Grey}}, \bibinfo {author} {\bibfnamefont {B.}~\bibnamefont
  {Dunn}},\ and\ \bibinfo {author} {\bibfnamefont {P.}~\bibnamefont {Simon}},\
  }\href@noop {} {\bibfield  {journal} {\bibinfo  {journal} {Nat. Energy}\
  }\textbf {\bibinfo {volume} {1}},\ \bibinfo {pages} {16070} (\bibinfo {year}
  {2016})}
  \BibitemShut {NoStop}%
\bibitem [{\citenamefont {Brogioli}(2009)}]{brogioli2009extracting}%
  \BibitemOpen
  \bibfield  {author} {\bibinfo {author} {\bibfnamefont {D.}~\bibnamefont
  {Brogioli}},\ }\href@noop {} {\bibfield  {journal} {\bibinfo  {journal}
  {Phys. Rev. Lett.}\ }\textbf {\bibinfo {volume} {103}},\ \bibinfo
  {pages} {058501} (\bibinfo {year} {2009})}\BibitemShut {NoStop}%
\bibitem [{\citenamefont {Janssen}\ \emph {et~al.}(2014)\citenamefont
  {Janssen}, \citenamefont {H{\"a}rtel},\ and\ \citenamefont {van
  Roij}}]{Janssen:2014aa}%
  \BibitemOpen
  \bibfield  {author} {\bibinfo {author} {\bibfnamefont {M.}~\bibnamefont
  {Janssen}}, \bibinfo {author} {\bibfnamefont {A.}~\bibnamefont {H{\"a}rtel}},\ and\ \bibinfo {author} {\bibfnamefont {R.}~\bibnamefont {van Roij}},\
  }\href {\doibase 10.1103/PhysRevLett.113.268501} {\bibfield  {journal}
  {\bibinfo  {journal} {Phys. Rev. Lett.}\ }\textbf {\bibinfo {volume}
  {113}},\ \bibinfo {pages} {268501} (\bibinfo {year} {2014})}\BibitemShut
  {NoStop}%
\bibitem [{\citenamefont {H{\"a}rtel}\ \emph
  {et~al.}(2015{\natexlab{a}})\citenamefont {H{\"a}rtel}, \citenamefont
  {Janssen}, \citenamefont {Weingarth}, \citenamefont {Presser},\ and\
  \citenamefont {van Roij}}]{hartel2015heat}%
  \BibitemOpen
  \bibfield  {author} {\bibinfo {author} {\bibfnamefont {A.}~\bibnamefont
  {H{\"a}rtel}}, \bibinfo {author} {\bibfnamefont {M.}~\bibnamefont {Janssen}},
  \bibinfo {author} {\bibfnamefont {D.}~\bibnamefont {Weingarth}}, \bibinfo
  {author} {\bibfnamefont {V.}~\bibnamefont {Presser}},\ and\ \bibinfo
  {author} {\bibfnamefont {R.}~\bibnamefont {van Roij}},\ }\href@noop {}
  {\bibfield  {journal} {\bibinfo  {journal} {Energy Environ. Sci.}\
  }\textbf {\bibinfo {volume} {8}},\ \bibinfo {pages} {2396} (\bibinfo {year}
  {2015}{\natexlab{a}})}\BibitemShut {NoStop}%
\bibitem [{\citenamefont {Hamelers}\ \emph {et~al.}(2013)\citenamefont
  {Hamelers}, \citenamefont {Schaetzle}, \citenamefont {Paz-Garc{\'\i}a},
  \citenamefont {Biesheuvel},\ and\ \citenamefont
  {Buisman}}]{hamelers2013harvesting}%
  \BibitemOpen
  \bibfield  {author} {\bibinfo {author} {\bibfnamefont {H.~V.~M.}~\bibnamefont
  {Hamelers}}, \bibinfo {author} {\bibfnamefont {O.}~\bibnamefont {Schaetzle}},
  \bibinfo {author} {\bibfnamefont {J.~M.}~\bibnamefont {Paz-Garc{\'\i}a}},
  \bibinfo {author} {\bibfnamefont {P.~M.}~\bibnamefont {Biesheuvel}},\ and\
  \bibinfo {author} {\bibfnamefont {C.~J.~N.}~\bibnamefont {Buisman}},\ }\href@noop
  {} {\bibfield  {journal} {\bibinfo  {journal} {Environ. Sci. Technol. Lett.}\ }
  \textbf{\bibinfo {volume} {1}},\ \bibinfo {pages} {31}
  (\bibinfo {year} {2014})}\BibitemShut {NoStop}%
\bibitem [{\citenamefont {Suss}\ \emph {et~al.}(2015)\citenamefont {Suss},
  \citenamefont {Porada}, \citenamefont {Sun}, \citenamefont {Biesheuvel},
  \citenamefont {Yoon},\ and\ \citenamefont {Presser}}]{suss2015water}%
  \BibitemOpen
  \bibfield  {author} {\bibinfo {author} {\bibfnamefont {M.~E.}~\bibnamefont
  {Suss}}, \bibinfo {author} {\bibfnamefont {S.}~\bibnamefont {Porada}},
  \bibinfo {author} {\bibfnamefont {X.}~\bibnamefont {Sun}}, \bibinfo {author}
  {\bibfnamefont {P.~M.}~\bibnamefont {Biesheuvel}}, \bibinfo {author}
  {\bibfnamefont {J.}~\bibnamefont {Yoon}},\ and\ \bibinfo {author}
  {\bibfnamefont {V.}~\bibnamefont {Presser}},\ }\href@noop {} {\bibfield
  {journal} {\bibinfo  {journal} {Energy Environ. Sci.}\ }\textbf
  {\bibinfo {volume} {8}},\ \bibinfo {pages} {2296} (\bibinfo {year}
  {2015})}\BibitemShut {NoStop}%
\bibitem [{\citenamefont {Xiong}\ \emph {et~al.}(2015)\citenamefont {Xiong},
  \citenamefont {Kundu},\ and\ \citenamefont {Fisher}}]{xiong2015thermal}%
  \BibitemOpen
  \bibfield  {author} {\bibinfo {author} {\bibfnamefont {G.}~\bibnamefont
  {Xiong}}, \bibinfo {author} {\bibfnamefont {A.}~\bibnamefont {Kundu}},\ and\
  \bibinfo {author} {\bibfnamefont {T.~S.}\ \bibnamefont {Fisher}},\
  }\href@noop {} {\emph {\bibinfo {title} {Thermal Effects in
  Supercapacitors}}}\ (\bibinfo  {publisher} {Springer, New York},\ \bibinfo {year}
  {2015})\BibitemShut {NoStop}%
\bibitem [{\citenamefont {Miller}(2006)}]{miller2006electrochemical}%
  \BibitemOpen
  \bibfield  {author} {\bibinfo {author} {\bibfnamefont {J.~R.}\ \bibnamefont
  {Miller}},\ }\href@noop {} {\bibfield  {journal} {\bibinfo  {journal}
  {Electrochim. Acta}\ }\textbf {\bibinfo {volume} {52}},\ \bibinfo {pages}
  {1703} (\bibinfo {year} {2006})}\BibitemShut {NoStop}%
\bibitem [{\citenamefont {Zhang}\ \emph {et~al.}(2016)\citenamefont {Zhang},
  \citenamefont {Wang}, \citenamefont {Lu}, \citenamefont {Hua},\ and\
  \citenamefont {Heng}}]{zhang2016reversible}%
  \BibitemOpen
  \bibfield  {author} {\bibinfo {author} {\bibfnamefont {X.}~\bibnamefont
  {Zhang}}, \bibinfo {author} {\bibfnamefont {W.}~\bibnamefont {Wang}},
  \bibinfo {author} {\bibfnamefont {J.}~\bibnamefont {Lu}}, \bibinfo {author}
  {\bibfnamefont {L.}~\bibnamefont {Hua}},\ and\ \bibinfo {author}
  {\bibfnamefont {J.}~\bibnamefont {Heng}},\ }\href@noop {} {\bibfield
  {journal} {\bibinfo  {journal} {Thermochim. Acta}\ }\textbf {\bibinfo
  {volume} {636}},\ \bibinfo {pages} {1} (\bibinfo {year} {2016})}\BibitemShut
  {NoStop}%
\bibitem [{\citenamefont {Pascot}\ \emph {et~al.}(2010)\citenamefont {Pascot},
  \citenamefont {Dandeville}, \citenamefont {Scudeller}, \citenamefont
  {Guillemet},\ and\ \citenamefont {Brousse}}]{pascot2010calorimetric}%
  \BibitemOpen
  \bibfield  {author} {\bibinfo {author} {\bibfnamefont {C.}~\bibnamefont
  {Pascot}}, \bibinfo {author} {\bibfnamefont {Y.}~\bibnamefont {Dandeville}},
  \bibinfo {author} {\bibfnamefont {Y.}~\bibnamefont {Scudeller}}, \bibinfo
  {author} {\bibfnamefont {Ph.}~\bibnamefont {Guillemet}},\ and\ \bibinfo
  {author} {\bibfnamefont {Th.}~\bibnamefont {Brousse}},\ }\href@noop {}
  {\bibfield  {journal} {\bibinfo  {journal} {Thermochim. Acta}\ }\textbf
  {\bibinfo {volume} {510}},\ \bibinfo {pages} {53} (\bibinfo {year}
  {2010})};
   \bibfield  {author} {\bibinfo {author} {\bibfnamefont {Y.}~\bibnamefont
  {Dandeville}}, \bibinfo {author} {\bibfnamefont {Ph.}~\bibnamefont
  {Guillemet}}, \bibinfo {author} {\bibfnamefont {Y.}~\bibnamefont
  {Scudeller}}, \bibinfo {author} {\bibfnamefont {O.}~\bibnamefont {Crosnier}},
  \bibinfo {author} {\bibfnamefont {L.}~\bibnamefont {Athouel}},\ and\
  \bibinfo {author} {\bibfnamefont {Th.}~\bibnamefont {Brousse}},\ }\href@noop
  {} {\bibfield  {journal} {\bibinfo  {journal} \emph{ibid.}\ }\textbf
  {\bibinfo {volume} {526}},\ \bibinfo {pages} {1} (\bibinfo {year}
  {2011})}
 \BibitemShut {NoStop}%
\bibitem [{\citenamefont {Gualous}\ \emph {et~al.}(2011)\citenamefont
  {Gualous}, \citenamefont {Louahlia},\ and\ \citenamefont
  {Gallay}}]{gualous2011supercapacitor}%
  \BibitemOpen
  \bibfield  {author} {\bibinfo {author} {\bibfnamefont {H.}~\bibnamefont
  {Gualous}}, \bibinfo {author} {\bibfnamefont {H.}~\bibnamefont {Louahlia}}, \
  and\ \bibinfo {author} {\bibfnamefont {R.}~\bibnamefont {Gallay}},\
  }\href@noop {} {\bibfield  {journal} {\bibinfo  {journal} {IEEE Trans. Power Electron.}\ }\textbf {\bibinfo {volume} {26}},\ \bibinfo {pages}
  {3402} (\bibinfo {year} {2011})}\BibitemShut {NoStop}%
\bibitem [{\citenamefont {d'Entremont}\ and\ \citenamefont
  {Pilon}(2014{\natexlab{a}})}]{d2014first}%
  \BibitemOpen
  \bibfield  {author} {\bibinfo {author} {\bibfnamefont {A.~L.}~\bibnamefont
  {d'Entremont}}\ and\ \bibinfo {author} {\bibfnamefont {L.}~\bibnamefont
  {Pilon}},\ }\href@noop {} {\bibfield  {journal} {\bibinfo  {journal} {J. Power Sources}\ }\textbf {\bibinfo {volume} {246}},\ \bibinfo {pages}
  {887} (\bibinfo {year} {2014}{\natexlab{a}})}\BibitemShut {NoStop}%
\bibitem [{\citenamefont {Lee}\ \emph {et~al.}(2014)\citenamefont {Lee},
  \citenamefont {Yi}, \citenamefont {Kim}, \citenamefont {Shin}, \citenamefont
  {Min}, \citenamefont {Choi},\ and\ \citenamefont {Lee}}]{lee2014modeling}%
  \BibitemOpen
  \bibfield  {author} {\bibinfo {author} {\bibfnamefont {J.}~\bibnamefont
  {Lee}}, \bibinfo {author} {\bibfnamefont {J.}~\bibnamefont {Yi}}, \bibinfo
  {author} {\bibfnamefont {D.}~\bibnamefont {Kim}}, \bibinfo {author}
  {\bibfnamefont {C.~B.}\ \bibnamefont {Shin}}, \bibinfo {author}
  {\bibfnamefont {K.-S.}\ \bibnamefont {Min}}, \bibinfo {author} {\bibfnamefont
  {J.}~\bibnamefont {Choi}},\ and\ \bibinfo {author} {\bibfnamefont {H.-Y.}\
  \bibnamefont {Lee}},\ }\href@noop {} {\bibfield  {journal} {\bibinfo
  {journal} {Energies}\ }\textbf {\bibinfo {volume} {7}},\ \bibinfo {pages}
  {8264} (\bibinfo {year} {2014})}\BibitemShut {NoStop}%
\bibitem [{\citenamefont {d'Entremont}\ and\ \citenamefont
  {Pilon}(2014{\natexlab{b}})}]{d2014first2}%
  \BibitemOpen
  \bibfield  {author} {\bibinfo {author} {\bibfnamefont {A.~L.}~\bibnamefont
  {d'Entremont}}\ and\ \bibinfo {author} {\bibfnamefont {L.}~\bibnamefont
  {Pilon}},\ }\href@noop {} {\bibfield  {journal} {\bibinfo  {journal} {Appl. Therm. Eng.}\ }
  \textbf {\bibinfo {volume} {67}},\ \bibinfo {pages}
  {439} (\bibinfo {year} {2014}); }
{\bibfield  {journal} {\bibinfo  {journal}
  {Int. J. Heat Mass Transfer}\ }\textbf {\bibinfo
  {volume} {75}},\ \bibinfo {pages} {637} (\bibinfo {year}
  {2014}{\natexlab{b}})}{\bibfield  {journal} {\bibinfo  {journal}
  {; J. Power Sources}\ }\textbf {\bibinfo
  {volume} {273}},\ \bibinfo {pages} {196} (\bibinfo {year}
  {2015}{\natexlab{c}}); }{\bibfield  {journal} {\bibinfo  {journal}
  {\emph{ibid}}\ }\textbf {\bibinfo
  {volume} {335}},\ \bibinfo {pages} {172} (\bibinfo {year}
  {2016}{\natexlab{c}}); }
  \bibfield  {author} {\bibinfo {author} {\bibfnamefont {L.}~\bibnamefont
  {Pilon}}, \bibinfo {author} {\bibfnamefont {H.}~\bibnamefont {Wang}},\ and\
  \bibinfo {author} {\bibfnamefont {A.~L.}~\bibnamefont {d'Entremont}},\
  }\href@noop {} {\bibfield  {journal} {\bibinfo  {journal} {J. Electrochem. Soc.}\ }\textbf {\bibinfo {volume} {162}},\ \bibinfo
  {pages} {A5158} (\bibinfo {year} {2015})}\BibitemShut {NoStop}%
\bibitem [{\citenamefont {Kumar}\ \emph {et~al.}(2015)\citenamefont {Kumar},
  \citenamefont {Mahalik}, \citenamefont {Strelcov}, \citenamefont {Tselev},
  \citenamefont {Lokitz}, \citenamefont {Kalinin}, \citenamefont {Sumpter}
  \emph {et~al.}}]{kumar2015microscopic}%
  \BibitemOpen
  \bibfield  {author} {\bibinfo {author} {\bibfnamefont {R.}~\bibnamefont
  {Kumar}}, \bibinfo {author} {\bibfnamefont {J.~P.}\ \bibnamefont {Mahalik}},
  \bibinfo {author} {\bibfnamefont {E.}~\bibnamefont {Strelcov}}, \bibinfo
  {author} {\bibfnamefont {A.}~\bibnamefont {Tselev}}, \bibinfo {author}
  {\bibfnamefont {B.~S.}\ \bibnamefont {Lokitz}}, \bibinfo {author}
  {\bibfnamefont {S.~V.}~\bibnamefont {Kalinin}}, \bibinfo {author} {\bibfnamefont
  {B.~G.}\ \bibnamefont {Sumpter}},\ }\href@noop {} {\bibfield
   {journal} {\bibinfo  {journal} {arXiv:1503.09141}}
  }\BibitemShut {NoStop}%
\bibitem [{\citenamefont {Pecharsky}\ and\ \citenamefont
  {Gschneidner~Jr}(1999)}]{pecharsky1999magnetocaloric}%
  \BibitemOpen
  \bibfield  {author} {\bibinfo {author} {\bibfnamefont {V.~K.}\ \bibnamefont
  {Pecharsky}}\ and\ \bibinfo {author} {\bibfnamefont {K.~A.}\ \bibnamefont
  {Gschneidner~Jr.}},\ }\href@noop {} {\bibfield  {journal} {\bibinfo  {journal}
  {J. Magn. Magn. Mater.}\ }\textbf {\bibinfo {volume}
  {200}},\ \bibinfo {pages} {44} (\bibinfo {year} {1999})}\BibitemShut
  {NoStop}%
\bibitem [{\citenamefont {Mischenko}\ \emph {et~al.}(2006)\citenamefont
  {Mischenko}, \citenamefont {Zhang}, \citenamefont {Scott}, \citenamefont
  {Whatmore},\ and\ \citenamefont {Mathur}}]{mischenko2006giant}%
  \BibitemOpen
  \bibfield  {author} {\bibinfo {author} {\bibfnamefont {A.~S.}~\bibnamefont
  {Mischenko}}, \bibinfo {author} {\bibfnamefont {Q.}~\bibnamefont {Zhang}},
  \bibinfo {author} {\bibfnamefont {J.~F.}~\bibnamefont {Scott}}, \bibinfo
  {author} {\bibfnamefont {R.~W.}~\bibnamefont {Whatmore}},\ and\ \bibinfo
  {author} {\bibfnamefont {N.~D.}~\bibnamefont {Mathur}},\ }\href@noop {}
  {\bibfield  {journal} {\bibinfo  {journal} {Science}\ }\textbf {\bibinfo
  {volume} {311}},\ \bibinfo {pages} {1270} (\bibinfo {year}
  {2006})}\BibitemShut {NoStop}%
\bibitem [{\citenamefont {Schweizer}\ and\ \citenamefont
  {Wauer}(2001)}]{schweizer2001atomistic}%
  \BibitemOpen
  \bibfield  {author} {\bibinfo {author} {\bibfnamefont {B.}~\bibnamefont
  {Schweizer}}\ and\ \bibinfo {author} {\bibfnamefont {J.}~\bibnamefont
  {Wauer}},\ }\href@noop {} {\bibfield  {journal} {\bibinfo  {journal} {Eur. Phys. J. B Condens. Matter}\ }\textbf
  {\bibinfo {volume} {23}},\ \bibinfo {pages} {383} (\bibinfo {year}
  {2001})}\BibitemShut {NoStop}%
\bibitem [{\citenamefont {Verwey}\ and\ \citenamefont
  {Overbeek}(1948)}]{verwey1948theory}%
  \BibitemOpen
  \bibfield  {author} {\bibinfo {author} {\bibfnamefont {E.~J.~W.}~\bibnamefont
  {Verwey}}\ and\ \bibinfo {author} {\bibfnamefont {J.~Th.~G.}~\bibnamefont
  {Overbeek}},\ }\href@noop {} {\emph {\bibinfo {title} {Theory of the
  Stability of Lyophobic Colloids
  }}}\ (\bibinfo  {publisher} {Elsevier, New Work},\ \bibinfo {year}
  {1948}), Chap. 3, p. 59\BibitemShut {NoStop}%
\bibitem [{\citenamefont {Groot}\ and\ \citenamefont
  {Mazur}(1962)}]{groot1962non}%
  \BibitemOpen
  \bibfield  {author} {\bibinfo {author} {\bibfnamefont {S.~R.~de}\ \bibnamefont
  {Groot}}\ and\ \bibinfo {author} {\bibfnamefont {P.}~\bibnamefont {Mazur}},\
  }\href@noop {} {\emph {\bibinfo {title} {Non-Equilibrium Thermodynamics}}}\
  (\bibinfo  {publisher} {North-Holland, Amsterdam},\ \bibinfo {year} {1962})\BibitemShut
  {NoStop}%
\bibitem [{\citenamefont {Haase}(1968)}]{haase1968thermodynamics}%
  \BibitemOpen
  \bibfield  {author} {\bibinfo {author} {\bibfnamefont {R.}~\bibnamefont
  {Haase}},\ }\href@noop {} {\emph {\bibinfo {title} {Thermodynamics of
  Irreversible Processes}}}\ (\bibinfo  {publisher} {Addison-Wesley, Reading, MA},\ \bibinfo
  {year} {1968})\BibitemShut {NoStop}%
\bibitem [{\citenamefont {Kontturi}\ \emph {et~al.}(2008)\citenamefont
  {Kontturi}, \citenamefont {Murtom{\"a}ki},\ and\ \citenamefont
  {Manzanares}}]{kontturi2008ionic}%
  \BibitemOpen
  \bibfield  {author} {\bibinfo {author} {\bibfnamefont {K.}~\bibnamefont
  {Kontturi}}, \bibinfo {author} {\bibfnamefont {L.}~\bibnamefont
  {Murtom{\"a}ki}},\ and\ \bibinfo {author} {\bibfnamefont {J.~A.}\
  \bibnamefont {Manzanares}},\ }\href@noop {} {\emph {\bibinfo {title} {Ionic
  Transport Processes: In Electrochemistry and Membrane Science}}}\ (\bibinfo  {publisher} {Oxford University Press,
  Oxford},\ \bibinfo {year} {2008})\BibitemShut {NoStop}%
\bibitem [{\citenamefont {Biesheuvel}\ \emph {et~al.}(2014)\citenamefont
  {Biesheuvel}, \citenamefont {Brogioli},\ and\ \citenamefont
  {Hamelers}}]{biesheuvel2014negative}%
  \BibitemOpen
  \bibfield  {author} {\bibinfo {author} {\bibfnamefont {P.~M.}~\bibnamefont
  {Biesheuvel}}, \bibinfo {author} {\bibfnamefont {D.}~\bibnamefont
  {Brogioli}},\ and\ \bibinfo {author} {\bibfnamefont {H.~V.~M.}~\bibnamefont
  {Hamelers}},\ }\href@noop {} {\bibfield  {journal} {\bibinfo  {journal}
  {arXiv:1402.1448}\ }}\BibitemShut
  {NoStop}%
\bibitem [{\citenamefont {Boon}\ and\ \citenamefont {van
  Roij}(2011)}]{boon2011blue}%
  \BibitemOpen
  \bibfield  {author} {\bibinfo {author} {\bibfnamefont {N.}~\bibnamefont
  {Boon}}\ and\ \bibinfo {author} {\bibfnamefont {R.}~\bibnamefont {van
  Roij}},\ }\href@noop {} {\bibfield  {journal} {\bibinfo  {journal} {Mol.
  Phys.}\ }\textbf {\bibinfo {volume} {109}},\ \bibinfo {pages} {1229}
  (\bibinfo {year} {2011})}\BibitemShut {NoStop}%
\bibitem [{\citenamefont {Boda}\ \emph {et~al.}(1999)\citenamefont {Boda},
  \citenamefont {Henderson}, \citenamefont {Chan},\ and\ \citenamefont
  {Wasan}}]{boda1999low}%
  \BibitemOpen
  \bibfield  {author} {\bibinfo {author} {\bibfnamefont {D.}~\bibnamefont
  {Boda}}, \bibinfo {author} {\bibfnamefont {D.}~\bibnamefont {Henderson}},
  \bibinfo {author} {\bibfnamefont {K.-Y.}\ \bibnamefont {Chan}},\ and\
  \bibinfo {author} {\bibfnamefont {D.~T.}\ \bibnamefont {Wasan}},\ }\href@noop
  {} {\bibfield  {journal} {\bibinfo  {journal} {Chem. Phys. Lett.}\
  }\textbf {\bibinfo {volume} {308}},\ \bibinfo {pages} {473} (\bibinfo {year}
  {1999})}\BibitemShut {NoStop}%
\bibitem [{\citenamefont {Reszko-Zygmunt}\ \emph {et~al.}(2005)\citenamefont
  {Reszko-Zygmunt}, \citenamefont {Soko{\l}owski}, \citenamefont {Henderson},\
  and\ \citenamefont {Boda}}]{reszko2005temperature}%
  \BibitemOpen
  \bibfield  {author} {\bibinfo {author} {\bibfnamefont {J.}~\bibnamefont
  {Reszko-Zygmunt}}, \bibinfo {author} {\bibfnamefont {S.}~\bibnamefont
  {Soko{\l}owski}}, \bibinfo {author} {\bibfnamefont {D.}~\bibnamefont
  {Henderson}},\ and\ \bibinfo {author} {\bibfnamefont {D.}~\bibnamefont
  {Boda}},\ }\href@noop {} {\bibfield  {journal} {\bibinfo  {journal} {J. Chem. Phys.}\ }\textbf {\bibinfo {volume} {122}},\ \bibinfo
  {pages} {084504} (\bibinfo {year} {2005})}\BibitemShut {NoStop}%
\bibitem [{\citenamefont {Jiang}\ \emph {et~al.}(2014)\citenamefont {Jiang},
  \citenamefont {Cao}, \citenamefont {Henderson},\ and\ \citenamefont
  {Wu}}]{jiang2014revisiting}%
  \BibitemOpen
  \bibfield  {author} {\bibinfo {author} {\bibfnamefont {J.}~\bibnamefont
  {Jiang}}, \bibinfo {author} {\bibfnamefont {D.}~\bibnamefont {Cao}}, \bibinfo
  {author} {\bibfnamefont {D.}~\bibnamefont {Henderson}},\ and\ \bibinfo
  {author} {\bibfnamefont {J.}~\bibnamefont {Wu}},\ }\href@noop {} {\bibfield
  {journal} {\bibinfo  {journal} {J. Chem. Phys.}\ }\textbf
  {\bibinfo {volume} {140}},\ \bibinfo {pages} {044714} (\bibinfo {year}
  {2014})}\BibitemShut {NoStop}%
\bibitem [{\citenamefont {Borukhov}\ \emph {et~al.}(1997)\citenamefont
  {Borukhov}, \citenamefont {Andelman},\ and\ \citenamefont
  {Orland}}]{borukhov1997steric}%
  \BibitemOpen
  \bibfield  {author} {\bibinfo {author} {\bibfnamefont {I.}~\bibnamefont
  {Borukhov}}, \bibinfo {author} {\bibfnamefont {D.}~\bibnamefont {Andelman}},\ and\ \bibinfo {author} {\bibfnamefont {H.}~\bibnamefont {Orland}},\
  }\href@noop {} {\bibfield  {journal} {\bibinfo  {journal} {Phys. Rev.
  Lett.}\ }\textbf {\bibinfo {volume} {79}},\ \bibinfo {pages} {435}
  (\bibinfo {year} {1997})}\BibitemShut {NoStop}%
\bibitem [{\citenamefont {Bikerman}(1942)}]{bikermanphil}%
  \BibitemOpen
  \bibfield  {author} {\bibinfo {author} {\bibfnamefont {J.~J.}~\bibnamefont
  {Bikerman}},\ }\href {\doibase 10.1080/14786444208520813} {\bibfield
  {journal} {\bibinfo  {journal} {Philos. Mag.}\ }\textbf
  {\bibinfo {volume} {33}},\ \bibinfo {pages} {384} (\bibinfo {year}
  {1942})};
  \bibfield  {author} {\bibinfo {author} {\bibfnamefont {V.}~\bibnamefont
  {Freise}},\ }\href@noop {} {\bibfield  {journal} {\bibinfo  {journal}
  {Z. Elektrochem.}\ }\textbf {\bibinfo {volume} {56}},\ \bibinfo {pages}
  {822} (\bibinfo {year} {1952})};
 \bibfield  {author} {\bibinfo {author} {\bibfnamefont {A.~A.}\ \bibnamefont
  {Kornyshev}},\ }\href@noop {} {\bibfield  {journal} {\bibinfo  {journal} {J. Phys. Chem. B}\ }\textbf {\bibinfo {volume} {111}},\
  \bibinfo {pages} {5545} (\bibinfo {year} {2007})}  
  \BibitemShut {NoStop}%
\bibitem [{\citenamefont {Kilic}\ \emph {et~al.}(2007)\citenamefont {Kilic},
  \citenamefont {Bazant},\ and\ \citenamefont {Ajdari}}]{kilic2007stericII}%
  \BibitemOpen
  \bibfield  {author} {\bibinfo {author} {\bibfnamefont {M.~S.}\ \bibnamefont
  {Kilic}}, \bibinfo {author} {\bibfnamefont {M.~Z.}\ \bibnamefont {Bazant}}, \
  and\ \bibinfo {author} {\bibfnamefont {A.}~\bibnamefont {Ajdari}},\
  }\href@noop {} {\bibfield  {journal} {\bibinfo  {journal} {Phys. Rev.
  E}\ }\textbf {\bibinfo {volume} {75}},\ \bibinfo {pages} {021503} (\bibinfo
  {year} {2007})}\BibitemShut {NoStop}%
%
\bibitem [{\citenamefont
  {Nightingale~Jr}(1959)}]{nightingale1959phenomenological}%
  \BibitemOpen
  \bibfield  {author} {\bibinfo {author} {\bibfnamefont {E.}~\bibnamefont
  {Nightingale~Jr.}},\ }\href@noop {} {\bibfield  {journal} {\bibinfo  {journal}
  {J. Phys. Chem. }\ }\textbf {\bibinfo {volume} {63}},\
  \bibinfo {pages} {1381} (\bibinfo {year} {1959})}\BibitemShut {NoStop}%
\bibitem [{\citenamefont {Kralj-Igli{\v{c}}}\ and\ \citenamefont
  {Igli{\v{c}}}(1996)}]{kralj1996simple}%
  \BibitemOpen
  \bibfield  {author} {\bibinfo {author} {\bibfnamefont {V.}~\bibnamefont
  {Kralj-Igli{\v{c}}}}\ and\ \bibinfo {author} {\bibfnamefont {A.}~\bibnamefont
  {Igli{\v{c}}}},\ }\href@noop {} {\bibfield  {journal} {\bibinfo  {journal}
  {J. Phys. II (France)}\ }\textbf {\bibinfo {volume} {6}},\ \bibinfo {pages} 
  {477} (\bibinfo {year} {1996})}; 
  \bibfield  {author} {\bibinfo {author} {\bibfnamefont {P.~M.}~\bibnamefont
  {Biesheuvel}}\ and\ \bibinfo {author} {\bibfnamefont {M.}~\bibnamefont
  {van~Soestbergen}},\ }\href@noop {} {\bibfield  {journal} {\bibinfo
  {journal} {J. Colloid Interface Sci.}\ }\textbf {\bibinfo 
  {volume} {316}},\ \bibinfo {pages} {490} (\bibinfo {year}
  {2007})}\BibitemShut {NoStop}%
\bibitem [{\citenamefont {Schmidt}\ and\ \citenamefont
  {Brader}(2013)}]{schmidt2013power}%
  \BibitemOpen
  \bibfield  {author} {\bibinfo {author} {\bibfnamefont {M.}~\bibnamefont
  {Schmidt}}\ and\ \bibinfo {author} {\bibfnamefont {J.~M.}\ \bibnamefont
  {Brader}},\ }\href@noop {} {\bibfield  {journal} {\bibinfo  {journal} {J. Chem. Phys.}\ }\textbf {\bibinfo {volume} {138}},\ \bibinfo
  {pages} {214101} (\bibinfo {year} {2013})}\BibitemShut {NoStop}%
\bibitem [{\citenamefont {Schmidt}(2011)}]{schmidt2011statics}%
  \BibitemOpen
  \bibfield  {author} {\bibinfo {author} {\bibfnamefont {M.}~\bibnamefont
  {Schmidt}},\ }\href@noop {} {\bibfield  {journal} {\bibinfo  {journal}
  {Phys. Rev. E}\ }\textbf {\bibinfo {volume} {84}},\ \bibinfo {pages}
  {051203} (\bibinfo {year} {2011})};
  \bibfield  {author} {\bibinfo {author} {\bibfnamefont {J.~G.}\ \bibnamefont
  {Anero}}, \bibinfo {author} {\bibfnamefont {P.}~\bibnamefont {Espa{\~n}ol}},
  \ and\ \bibinfo {author} {\bibfnamefont {P.}~\bibnamefont {Tarazona}},\
  }\href@noop {} {\bibfield  {journal} {\bibinfo  {journal} {J. Chem. Phys.}\ }\textbf {\bibinfo {volume} {139}},\ \bibinfo {pages}
  {034106} (\bibinfo {year} {2013})}  
  \BibitemShut {NoStop}%
\bibitem [{\citenamefont {Lee}\ \emph {et~al.}(2015)\citenamefont {Lee},
  \citenamefont {Kondrat}, \citenamefont {Vella},\ and\ \citenamefont
  {Goriely}}]{kondrat2015dynamics}%
  \BibitemOpen
  \bibfield  {author} {\bibinfo {author} {\bibfnamefont {A.~A.}\ \bibnamefont
  {Lee}}, \bibinfo {author} {\bibfnamefont {S.}~\bibnamefont {Kondrat}},
  \bibinfo {author} {\bibfnamefont {D.}~\bibnamefont {Vella}},\ and\ \bibinfo
  {author} {\bibfnamefont {A.}~\bibnamefont {Goriely}},\ }\href@noop {}
  {\bibfield  {journal} {\bibinfo  {journal} {Phys. Rev. Lett.}\
  }\textbf {\bibinfo {volume} {115}},\ \bibinfo {pages} {106101} (\bibinfo
  {year} {2015})}\BibitemShut {NoStop}%
\bibitem [{\citenamefont {Hatlo}\ \emph {et~al.}(2012)\citenamefont {Hatlo},
  \citenamefont {van~Roij},\ and\ \citenamefont {Lue}}]{hatlo2012electric}%
  \BibitemOpen
  \bibfield  {author} {\bibinfo {author} {\bibfnamefont {M.~M.}~\bibnamefont
  {Hatlo}}, \bibinfo {author} {\bibfnamefont {R.}~\bibnamefont {van~Roij}}, \
  and\ \bibinfo {author} {\bibfnamefont {L.}~\bibnamefont {Lue}},\ }\href@noop
  {} {\bibfield  {journal} {\bibinfo  {journal} {Europhys. Lett.}\
  }\textbf {\bibinfo {volume} {97}},\ \bibinfo {pages} {28010} (\bibinfo {year}
  {2012})}\BibitemShut {NoStop}%
\bibitem [{\citenamefont {H{\"a}rtel}\ \emph
  {et~al.}(2015{\natexlab{b}})\citenamefont {H{\"a}rtel}, \citenamefont
  {Janssen}, \citenamefont {Samin},\ and\ \citenamefont {van
  Roij}}]{hartel2015fundamental}%
  \BibitemOpen
  \bibfield  {author} {\bibinfo {author} {\bibfnamefont {A.}~\bibnamefont
  {H{\"a}rtel}}, \bibinfo {author} {\bibfnamefont {M.}~\bibnamefont {Janssen}},
  \bibinfo {author} {\bibfnamefont {S.}~\bibnamefont {Samin}},\ and\ \bibinfo
  {author} {\bibfnamefont {R.}~\bibnamefont {van Roij}},\ }\href@noop {}
  {\bibfield  {journal} {\bibinfo  {journal} {J. Phys. Condens. Matter}\ }
  \textbf {\bibinfo {volume} {27}},\ \bibinfo {pages} {194129}
  (\bibinfo {year} {2015}{\natexlab{b}})}\BibitemShut {NoStop}%
\bibitem [{\citenamefont {Chmiola}\ \emph {et~al.}(2006)\citenamefont
  {Chmiola}, \citenamefont {Yushin}, \citenamefont {Gogotsi}, \citenamefont
  {Portet}, \citenamefont {Simon},\ and\ \citenamefont
  {Taberna}}]{chmiola2006anomalous}%
  \BibitemOpen
  \bibfield  {author} {\bibinfo {author} {\bibfnamefont {J.}~\bibnamefont
  {Chmiola}}, \bibinfo {author} {\bibfnamefont {G.}~\bibnamefont {Yushin}},
  \bibinfo {author} {\bibfnamefont {Y.}~\bibnamefont {Gogotsi}}, \bibinfo
  {author} {\bibfnamefont {C.}~\bibnamefont {Portet}}, \bibinfo {author}
  {\bibfnamefont {P.}~\bibnamefont {Simon}},\ and\ \bibinfo {author}
  {\bibfnamefont {P.-L.}\ \bibnamefont {Taberna}},\ }\href@noop {} {\bibfield
  {journal} {\bibinfo  {journal} {Science}\ }\textbf {\bibinfo {volume}
  {313}},\ \bibinfo {pages} {1760} (\bibinfo {year} {2006})}; 
  \bibfield  {author} {\bibinfo {author} {\bibfnamefont {M.~S.}~\bibnamefont
  {Loth}}, \bibinfo {author} {\bibfnamefont {B.}~\bibnamefont {Skinner}}, \
  and\ \bibinfo {author} {\bibfnamefont {B.~I.}~\bibnamefont {Shklovskii}},\
  }\href@noop {} {\bibfield  {journal} {\bibinfo  {journal} {Phys. Rev.
  E}\ }\textbf {\bibinfo {volume} {82}},\ \bibinfo {pages} {016107} (\bibinfo
  {year} {2010}{\natexlab{a}})}; 
 {\bibfield  {journal} 
  \textbf {\bibinfo {volume} {82}},\ \bibinfo {pages} {056102} (\bibinfo
  {year} {2010}{\natexlab{b}})}; 
  \bibfield  {author} {\bibinfo {author} {\bibfnamefont {C.}~\bibnamefont
  {Merlet}}, \bibinfo {author} {\bibfnamefont {D.~T.}\ \bibnamefont {Limmer}},
  \bibinfo {author} {\bibfnamefont {M.}~\bibnamefont {Salanne}}, \bibinfo
  {author} {\bibfnamefont {R.}~\bibnamefont {van~Roij}}, \bibinfo {author}
  {\bibfnamefont {P.~A.}\ \bibnamefont {Madden}}, \bibinfo {author}
  {\bibfnamefont {D.}~\bibnamefont {Chandler}},\ and\ \bibinfo {author}
  {\bibfnamefont {B.}~\bibnamefont {Rotenberg}},\ }\href@noop {} {\bibfield
  {journal} {\bibinfo  {journal} {J. Phys. Chem. C}\ 
  }\textbf {\bibinfo {volume} {118}},\ \bibinfo {pages} {18291} (\bibinfo
  {year} {2014})}; 
  \bibfield  {author} {\bibinfo {author} {\bibfnamefont {D.~T.}\ \bibnamefont
  {Limmer}},\ }\href@noop {} {\bibfield  {journal} {\bibinfo  {journal}
  {Phys. Rev. Lett.}\ }\textbf {\bibinfo {volume} {115}},\ \bibinfo
  {pages} {256102} (\bibinfo {year} {2015})}\BibitemShut {NoStop}%
\bibitem [{\citenamefont {Curtiss}\ and\ \citenamefont
  {Bird}(1999)}]{curtiss1999multicomponent}%
  \BibitemOpen
  \bibfield  {author} {\bibinfo {author} {\bibfnamefont {C.~F.}~\bibnamefont
  {Curtiss}}\ and\ \bibinfo {author} {\bibfnamefont {R.~B.}\ \bibnamefont
  {Bird}},\ }\href@noop {} {\bibfield  {journal} {\bibinfo  {journal}
  {Ind. Eng. Chem. Res.}\ }\textbf {\bibinfo {volume}
  {38}},\ \bibinfo {pages} {2515} (\bibinfo {year} {1999})}\BibitemShut {NoStop}%
\bibitem [{\citenamefont {De~Groot}\ \emph {et~al.}(1953)\citenamefont
  {De~Groot}, \citenamefont {Mazur},\ and\ \citenamefont
  {Tolhoek}}]{de1953thermodynamics}%
  \BibitemOpen
  \bibfield  {author} {\bibinfo {author} {\bibfnamefont {S.~R.}~\bibnamefont
  {de~Groot}}, \bibinfo {author} {\bibfnamefont {P.}~\bibnamefont {Mazur}},\
  and\ \bibinfo {author} {\bibfnamefont {H.~A.}~\bibnamefont {Tolhoek}},\
  }\href@noop {} {\bibfield  {journal} {\bibinfo  {journal} {Physica (Amsterdam)}\ }\textbf
  {\bibinfo {volume} {19}},\ \bibinfo {pages} {549} (\bibinfo {year}
  {1953})}\BibitemShut {NoStop}%
\bibitem [{\citenamefont {Kell}(1975)}]{kell1975density}%
  \BibitemOpen
  \bibfield  {author} {\bibinfo {author} {\bibfnamefont {G.~S.}\ \bibnamefont
  {Kell}},\ }\href@noop {} {\bibfield  {journal} {\bibinfo  {journal} {J. Chem. Eng. Data}\ }\textbf {\bibinfo {volume} {20}},\
  \bibinfo {pages} {97} (\bibinfo {year} {1975})}\BibitemShut {NoStop}%
\bibitem [{\citenamefont {Davidson}(1962)}]{davidson1962}%
  \BibitemOpen
  \bibfield  {author} {\bibinfo {author} {\bibfnamefont {N.}~\bibnamefont
  {Davidson}},\ }\href@noop {} {\emph {\bibinfo {title} {Statistical
  Mechanics}}}\ (\bibinfo  {publisher} {Dover, New York},\ \bibinfo {year}
  {2003}), p. 514   \BibitemShut {NoStop}%
\bibitem [{\citenamefont {Bird}\ \emph {et~al.}(2007)\citenamefont {Bird},
  \citenamefont {Lightfoot},\ and\ \citenamefont
  {Stewart}}]{bird2007transport}%
  \BibitemOpen
  \bibfield  {author} {\bibinfo {author} {\bibfnamefont {R.~B.}\ \bibnamefont
  {Bird}}, \bibinfo {author} {\bibfnamefont {E.~W.}\ \bibnamefont {Stewart}},\ and\ \bibinfo {author} {\bibfnamefont {E.~N.}\ \bibnamefont {Lightfoot}},\
  }\href@noop {} {\emph {\bibinfo {title} {Transport Phenomenon}}}\ (\bibinfo
  {publisher} {Wiley, New York},\ \bibinfo {year} {2002}) \BibitemShut {NoStop}%
\bibitem [{\citenamefont {DeVoe}(2001)}]{devoe2001thermodynamics}%
  \BibitemOpen
  \bibfield  {author} {\bibinfo {author} {\bibfnamefont {H.}~\bibnamefont
  {DeVoe}},\ }\href@noop {} {\emph {\bibinfo {title} {Thermodynamics and
  Chemistry}}}\ (\bibinfo  {publisher} {Prentice Hall Upper Saddle River, NJ},\
  \bibinfo {year} {2001})\BibitemShut {NoStop}%
\end{thebibliography}

%

\end{document}